\documentclass[11pt]{article}
\usepackage[utf8]{inputenc}
\usepackage{fullpage}
\usepackage{graphicx}
\usepackage{subcaption}
\usepackage{amsmath}
\usepackage{amsfonts} 
\usepackage{algorithm}
\usepackage{mathtools}
\usepackage{mathrsfs}
\usepackage{cite}
\usepackage[colorlinks=true,urlcolor=blue]{hyperref}
\usepackage[noend]{algpseudocode}
\usepackage{bbm}

\usepackage{url}
\usepackage{amsmath,amsthm,amssymb,amsbsy}
\usepackage{mathdots}
\usepackage{paralist}
\usepackage{xcolor}
\usepackage{color}
\usepackage{graphicx}
\usepackage{algorithm,algpseudocode}
\usepackage{comment}

\usepackage{fancyhdr}
\usepackage{cite}
\usepackage{cleveref}
\usepackage{enumerate}

\theoremstyle{remark}

\newcommand{\vct}[1]{\boldsymbol{#1}}
\newcommand{\mtx}[1]{\boldsymbol{#1}}
\newcommand{\inner}[1]{\left<#1\right>}

\newcommand{\trace}{\operatorname{trace}}

\DeclareMathOperator*{\maximize}{\text{maximize}}
\DeclareMathOperator*{\argmin}{\text{arg~min}}

\newcommand{\eps}{\epsilon}

\newcommand{\vf}{\vct{f}}

\newcommand{\vs}{\vct{s}}

\newcommand{\vu}{\vct{u}}

\newcommand{\vx}{\vct{x}}
\newcommand{\vy}{\vct{y}}
\newcommand{\vz}{\vct{z}}
\newcommand{\valpha}{\vct{\alpha}}

\newcommand{\vphi}{\vct{\phi}}

\newcommand{\mA}{\mtx{A}}
\newcommand{\mB}{\mtx{B}}

\newcommand{\mE}{\mtx{E}}

\newcommand{\mN}{\mtx{N}}

\newcommand{\mR}{\mtx{R}}
\newcommand{\mS}{\mtx{S}}

\newcommand{\mX}{\mtx{X}}
\newcommand{\mY}{\mtx{Y}}

\newcommand{\mPsi}{\mtx{\Psi}}

\graphicspath{{./figs/}}

\newlength{\imgwidth}
\newboolean{twoColVersion}
\setboolean{twoColVersion}{false}
\newcommand{\twoCol}[2]{\ifthenelse{\boolean{twoColVersion}} {#1} {#2} }

\newcommand{\norm}[1]{\left\lVert#1\right\rVert}
\newcommand{\veta}{\vct{\eta}}
\newcommand{\vtheta}{\vct{\theta}}

\title{Iterative Broadband Source Localization}
\author{Coleman DeLude, Rakshith Sharma, Santhosh Karnik, Christopher Hood,\\ Mark Davenport, and Justin Romberg}
\date{October 20, 2022}

\begin{document}

\maketitle

\begin{abstract}
In this paper we consider the problem of localizing a set of broadband sources from a finite window of measurements. In the case of narrowband sources this can be reduced to the problem of spectral line estimation, where our goal is simply to estimate the active frequencies from a weighted mixture of pure sinusoids. There exists a plethora of modern and classical methods that effectively solve this problem. However, for a wide variety of applications the underlying sources are \emph{not} narrowband and can have an appreciable amount of bandwidth. In this work, we extend classical greedy algorithms for sparse recovery (e.g., orthogonal matching pursuit) to localize broadband sources. We leverage models for samples of broadband signals based on a union of \emph{Slepian subspaces}, which are more aptly suited for dealing with spectral leakage and dynamic range disparities. We show that by using these models, our adapted algorithms  can successfully localize broadband sources under a variety of adverse operating scenarios. Furthermore, we show that our algorithms outperform complementary methods that use more standard Fourier models. We also show that we can perform estimation from compressed measurements with little loss in fidelity as long as the number of measurements are on the order of the signal's implicit degrees of freedom. We conclude with an in-depth application of these ideas to the problem of localization in multi-sensor arrays.  
\end{abstract}
\section{Introduction}

At its core this paper revisits and extends the classical problem of source localization. Specifically, we assume that we observe the superposition of $L$ sources $x_1(t), \ldots, x_L(t)$ corrupted by noise $\eta(t)$:
\begin{align}
    \label{eq:source_loc_cont}
    y(t) = \sum_{\ell=1}^L x_\ell (t) + \eta(t).
\end{align}
We then sample $y(t)$ at $N$ points $\{t_n\}_{n=1}^N$ and denote the vector of these samples by $\vy$. Broadly speaking, source localization can be accomplished by identifying certain key parameters of the component signals from the observations $\vy$.
% \begin{align}
%     \label{eq:source_loc_samp}
%     \vy = \sum_{\ell=1}^L \vx_\ell  + \veta.
% \end{align}
Different applications will entail varying choices in the sampling domain and the specific structure imposed on the $x_\ell(t)$.  Perhaps the most commonly studied variation is when when $x_\ell(t) = c_\ell e^{j2\pi f_\ell t}$ for $c_\ell \in \mathbb{C},~f_\ell\in\mathbb{R}$ and we seek to estimate $\{(c_\ell,~f_\ell)\}_{\ell=1}^L$ from $\vy$, in which case this problem is called \emph{spectral line estimation} (SLE). SLE arises in a variety of source localization problems. For example, in \emph{direction-of-arrival} estimation (DOA), $\vy$ consists of a ``snapshot" from a multi-sensor array and we wish to estimate the angle at which each $x_\ell(t)$ is impinging on an array.\footnote{In this instance each $x_\ell(t)$ manifests as a plane-wave moving through space.} Classically, DOA is most often performed under a \emph{narrowband} %\footnote{``Narrowband" will be formally defined in Section~\ref{sec:spectral_line_est}.} 
assumption which makes it effectively equivalent to SLE. In fact, by leveraging a narrowband signal model a wide variety of important problems in communications, seismology, and radar can be cast as a SLE problem\cite{Stoica:2005}.

The narrowband signal model coupled with existing works on SLE form the foundation upon which a more general class of source localization algorithms can be built. For this reason Section~\ref{sec:spectral_line_est} offers a thorough overview of both recent and classical approaches to this problem. However, our goal in this paper is to explicitly diverge from this narrowband assumption and develop robust source localization algorithms that can accommodate \emph{broadband} sources. With even a relatively modest increase in the bandwidth of the sources, the narrowband assumption fails to hold, even approximately. As will be discussed in further detail in Section~\ref{sec:broadband_signal_estimation} this leads to a breakdown in the performance of existing localization algorithms. This forms the motivation for us to explore models and methods that more aptly suit broadband signals and can operate under a wider range of adverse conditions.

%To bolster our use case we define a condition under which the narrowband assumption fails, which occurs at a relatively low but non-zero bandwidth. 
In developing our algorithms, we make use of a specialized subspace model for broadband signals that is based on the \textit{Discrete Prolate Spheroidal Sequences} (DPSSs)\cite{SlepianV}. The finitely truncated DPSSs (when appropriately modulated), also known as \emph{Slepian basis} vectors, provide an optimal subspace representation for broadband signals sampled over a finite interval and form the backbone of our algorithms\cite{davenport2012compressive}. Our algorithms borrow from existing ``greedy" algorithms for sparse approximation. In an iterative fashion we alternate between projecting the signal (or residual) onto all possible Slepian subspaces and identifying the subspaces that capture the most energy to form an estimate of each component signals spectral support. Slepian subspaces are also used at each iteration to ``null'' the spectral components identified in previous iterations. The use of Slepian subspace models allows our algorithms to be robust to a large dynamic range between the signal powers when there are multiple active sources as well as operate at lower SNRs than traditional methods.
 
To demonstrate the utility of our proposed algorithms, we provide a host of experimental results. We first review existing narrowband source localization methods and show that they fail to generalize to the broadband case. Our proposed algorithms are then compared in a variety of scenarios. We then consider the problem of spectral support estimation from compressed samples. We show that when the number of samples is proportional to the inherent dimensionality (explained in detail in Section~\ref{sec:dictionary}) of the signal, our algorithms successfully identify the spectral support of each source. We then consider a multi-sensor array receiving several broadband signals from various DOAs. We apply our algorithms to this realistic scenario and demonstrate that they successfully identify the parameters associated with each source.

\begin{figure}[t]
        \centering
        \begin{subfigure}[b]{0.4\textwidth}
            \centering
            \includegraphics[width=\textwidth,height = .65\textwidth]{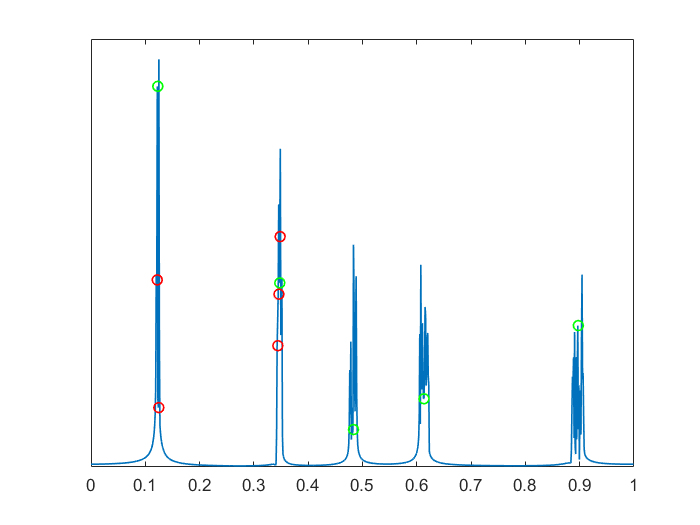}
            \caption[Broadband OMP vs SLE OMP]%
            {{\small}}    
            \label{fig:BroabandOMP_vs_SLEOMP}
        \end{subfigure}
        %\hfill
        \begin{subfigure}[b]{0.4\textwidth}  
            \centering 
            \includegraphics[width=\textwidth,height = .65\textwidth]{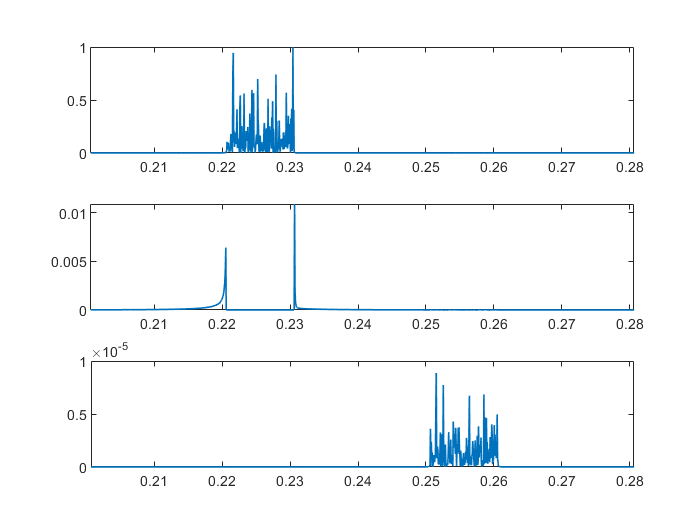}
            \caption[Nulling Step]%
            {{\small}}    
            \label{fig:Nulling_Step}
        \end{subfigure}
        \caption[Intro Figures]
        {\small \sl (a) Comparison of standard spectral line estimation based OMP frequency estimates (red circles) and broadband adapted OMP estimates (green circle). (b) For a two broadband signals with significant dynamic range disparities (top), we can null out sources using a Fourier dictionary projection (middle) or a Slepian dictionary projection (bottom).}
        \label{fig:Intro_Figures}
\end{figure}

  We provide here a brief snapshot of the nature of the experimental results given throughout the rest of the paper. In Figure~\ref{fig:Intro_Figures}(a), we demonstrate a specific instance of one of our proposed algorithms for broadband DOA estimation and compare it to the classical narrowband version of the same algorithm. While classical algorithms fail to identify all the sources due to the non-negligible bandwidth, our algorithm is able to correctly identify them. In Figure~\ref{fig:Intro_Figures}(b), we demonstrate the utility of the Slepian subspace model as compared to the traditional Fourier model for broadband signals. Specifically, we consider two active broadband signals, but with a vast difference in signal powers, with one signal being $10^5$ times stronger than the other. The traditional broadband Fourier model, in which a signal is represented by orthogonal discrete Fourier transform (DFT) basis vectors, fails to entirely capture the stronger source leading to repeated identification of the same source. Our Slepian signal model however is successful in identifying the full spectral support of the stronger source, hence leading to the correct identification of the other sources in subsequent iterations.
  
\section{Existing Approaches to Source Localization}
\label{sec:spectral_line_est}

\subsection{Localization via spectral line estimation}
We now briefly return to the previously described case of spectral line estimation (SLE). In the noiseless case, we observe the superposition of $L$ component sinusoidal tones:
\begin{align}
    \label{eq:sinus_SLE}
    s(t) & =\sum_{\ell=1}^L x_\ell(t)  =\sum_{\ell=1}^L c_\ell e^{j 2 \pi f_\ell t}
    \end{align} 
This is an idealized narrowband signal model. Here the Fourier transform of \eqref{eq:sinus_SLE} is a superposition of Dirac delta distributions $\widehat{S}(f)  = \sum_{\ell=1}^L c_\ell \delta(f - f_\ell)$ and the component signals have effectively zero bandwidth. 

If we had access to the entire temporal extent of the signal, then the SLE problem becomes trivial. Simple peak picking in the frequency domain would suffice to determine the spectral support while a least-squares problem could be used to determine the coefficients. However, we observe \eqref{eq:sinus_SLE} through a finite set of $N$ samples. We assume that these samples are taken uniformly over some interval with sampling frequency $f_s$ satisfying the Nyquist rate criterion (determined by the component with the largest frequency). A point of emphasis is that $f_\ell \in \mathbb{R}$ can generally be arbitrary, meaning that $f_\ell/f_s$ is likely not an integer multiple of $1/N$. These ``off-grid" frequencies subject the signal to spectral leakage during the sampling process wherein energy from the component signals bleed into one another, biasing the spectral peaks and amplitudes. A visualization of this phenomena is provided in Figure~\ref{fig:Coefficients}, where the presence of sidelobes about the sources is indicative of spectral leakage.
% \begin{figure}[h]
%     \centering
%     \includegraphics[scale=0.15]{figs/sig_spec.png}
%     \caption{\small \sl A single instance of the line spectral estimation problem with 5 active frequencies. The sinusoids are not necessarily chosen to be ``on-grid'' points, leading to spectral leakage.}
%     \label{fig:SLE_model}
% \end{figure}
The bias incurred from spectral leakage makes the SLE problem non-trivial, and we cannot determine the parameters of \eqref{eq:sinus_SLE} perfectly through standard methods of spectral estimation. This inconvenient fact has motivated an impressive body of work that spans more than two centuries of research.

The SLE problem has roots dating back to Prony's method \cite{Prony:1795}. Methods based on statistical signal models leveraging eigendecompositions of the signal covariance matrix were later pioneered in \cite{Caratheodory:1911} and subsequently rediscovered in \cite{Pisarenko:1973}. Initially considered to be less proficient at SLE than Prony's method \cite{Marple:1979}, this methodology would form the inspiration for the MUSIC algorithm \cite{Schmidt:1986}. Within a similar timeframe methods leveraging the rotational invariance of the signal subspace to directly estimate the component frequencies were developed to form the ESPRIT algorithm \cite{Roy:1989}. Accompanying these statistical methods, deterministic Prony-like methods such as the matrix pencil were later developed \cite{Hua:1988,Hua:1992}. The generalized eigenvalue problem solved in the matrix pencil method (with Prony's methods enveloped as a special case) is more stable to noise than the standard root finding process associated with Prony's method \cite{Hua:1990}. The matrix pencil, MUSIC, and ESPRIT algorithms form what we refer to as the ``classical" SLE methods.

More recently, advances in sparse approximation has led to the creation of a more optimization based perspective on SLE. Applying $\ell_1-\ell_2$ optimization to source localization, specifically in the context of arrays, was first presented in \cite{Fuchs:1996}. The work in \cite{Donoho:2005,Fuchs:2005} showed that when the components signals reside on on-grid frequencies and have positive weightings the signals can be perfectly recovered via $\ell_1$-minimization. Sparsity constraints were also shown to yield meaningful performance improvements in more practical radar based applications in \cite{Malioutov:2005}. 

Works such as \cite{DUARTE2013111,Eftekhari:2015,Fannjiang:2012} adapt compressed sensing recovery algorithms such as OMP, CoSaMP, subspace pursuit, and $\ell_1$-minimization to SLE when the observations in \eqref{eq:source_loc_cont} are viewed through a dimensionality reducing sensing matrix. The approach in each is similar, the component signals are assumed to admit a sparse representation in an overcomplete discrete Fourier dictionary. To overcome the dictionary's coherence the elements are split into coherent frames, and the sources can be coarsely localized to a subset of these frames. This approximation is then refined via a local optimization step that each author handles differently. In a similar manner to compressed sensing, the authors of~\cite{Chen:2014} proposed a 2-D SLE technique posed in a structured matrix completion framework. The process leverages the ``matrix enhancement" method developed in \cite{Hua:1992} to yield recovery guarantees in traditionally adverse scenarios. However, the paper is largely centered on the matrix completion aspect of the problem and defers the SLE portion to the methods proposed in \cite{Hua:1992,Hua:1990}.

A common theme amongst the above sparse approximation and compressed sensing SLE methods is the assumption that the signal is sparse in a finite dictionary. This precludes the scenario where the underlying frequencies of the component signals lie off-grid. As previously discussed, in the traditional SLE problem frequencies are permitted to lie on a \emph{continuum}. The use of over-complete dictionaries is meant to help circumvent this problem, but is ultimately avoiding the true nature of the issue at hand. 

TV/atomic norm minimization based methods\footnote{In the context of SLE, the total variation (TV) norm and atomic norm are essentially equivalent~\cite{Tang:2013}.} reconcile this issue, and are effectively a generalization of $\ell_1$-minimization to a continuous setting \cite{Chi20harness}. Utilizing TV norm minimization, the authors of~\cite{Castro:2011} were able to extend the work of \cite{Donoho:2005,Fuchs:2005} to operate on a continuum of frequencies (assuming a positive weighting of the sinusoids). Subsequently, in~\cite{Candes:2014} it was shown that, under a mild separation constraint, exact recovery of the component signal parameters could be achieved in the absence of noise via TV norm minimization. In~\cite{Candes:2013}, the sequel to \cite{Candes:2014}, strong theoretical guarantees on the accuracy of reconstruction from noisy measurements were established for the same atomic norm denoising framework presented in~\cite{Bhaskar:2013}. The work of~\cite{Fernandez:2013} decoupled the support and amplitude estimation errors and showed that the support estimation in atomic norm denoising is generally very accurate. Finally,~\cite{Tang:2013} showed that atomic norm minimization can be applied to compressed sensing in a manner similar to $\ell_1$-minimization to provide exact signal recovery. In summary TV/atomic norm based methods of SLE represent the state-of-the-art in terms of theoretical guarantees.

Alongside TV/atomic norm based methods, several authors have attempted to ``modernize" classical methods and revisit sparse recovery. The motivation for this being that such methods can offer computational savings. The work of~\cite{Aich:2017} provides empirical comparisons of OMP and CoSaMP algorithms for compressed DOA estimation. It has been shown both theoretically and empirically that single observation versions of MUSIC can be competitive with atomic norm based methods \cite{Liao:2014}, and have practical applications in realistic scenarios \cite{Maisto:2022}.

We note that an independent method based on finite \emph{rates of innovation} was shown to be able to recover Dirac trains from lowpass measurements in \cite{Vetterli:2002,Dragotti:2007}. Though some theoretical guarantees are presented, the method of recovery fundamentally relies on polynomial root finding. As a result, this approach is generally considered to be ill-suited for use in the presence of noise~\cite{Tan:2008}.

To accompany this review of SLE methods, Section~\ref{sec:numerical_SLE} offers a performance comparison of OMP, CoSaMP, $\ell_1$-minimization, and atomic norm minimization. Though it is readily apparent from these results that atomic norm minimization provides the best performance, a key observation is that the level of improvement is not particularly drastic. This is important when we consider how we will transition to the broadband regime since modifying most of these algorithms to account for bandwidth is generally difficult. In particular, for atomic norm minimization, it is not even clear how to define an atomic set for general broadband signals.\footnote{Even in the discrete case e.g. $\ell_1$-minimization we must carefully devise some notion of group sparsity.} Even if such a set existed it is not clear that it could be cast to a tractable framework \cite{Suliman:2021}. On the other hand, we will see that the iterative OMP and CoSaMP algorithms \emph{can} be modified to account for bandwidth in a fairly natural manner.

\subsection{Beyond SLE}
The problem of source localization in the broadband regime has been studied to a lesser extant than its narrowband counterpart. Some formulations of the standard atomic norm based frameworks allow for a known bandlimited point spread function (PSF) or kernel to be convolved with the component signals \cite{Chi20harness}. However, this does not generalize to arbitrary bandlimited signals and it is generally assumed that a de-convolution step has occurred prior to the optimization stage. The advances in~\cite{Wakin:2020,Suliman:2018} can be interpreted as an improvement upon this by attempting to estimate the sinusoidal components \emph{and} the kernel that they have been convovled with. Both methods hinge on said kernel lying in a known low dimensional subspace. As will be discussed in Section~\ref{sec:broadband_signal_estimation} this diverges from our general broadband signal model in which \emph{none} of the component signals lie exactly in a low dimensional subspace and merely reside close to one.

In terms of our approach and methodology, our work most closely resembles \cite{davenport2012compressive}, which utilizes a union of Slepian spaces model similar to our own. However, this work splits the spectrum into fixed intervals from which a dictionary can be defined. Signals are then determined to be present on these intervals or not, localizing the spectral support to a sub-band. While this accounts for bandwidth, like most of the applications of sparse approximation techniques to spectral estimation described above it again makes the flawed assumption that the component signals are centered on a (relatively coarse) grid of frequencies. While the proposed greedy algorithms of~\cite{davenport2012compressive} are similar to our own in the sense that they approach the problem from a signal space perspective \cite{Davenport:SSCoSaMP}, our algorithms distinguish themselves by not using a fixed dictionary. Instead we allow the Slepian spaces to lie on a continuum of possible intervals, and adaptively build a representative subspace.

\section{Slepian Representations for Broadband Signals}
\label{sec:broadband_signal_estimation}
%%%%%%%%%%%%%%%%%%%%%%%%%%%%%%%%%%%%%%%%%%%%%%%%%%%%%%%%%%%
\subsection{Broadband signal model}
Our broadband source localization algorithms hinge on a carefully developed subspace model that can produce low-dimensional representations of signals with appreciable bandwidth. This model can be motivated from a variety of perspectives, but perhaps the most natural is to assume a stochastic model on the underlying sources. Specifically, let us return to \eqref{eq:source_loc_cont} and assume that each $x_\ell(t)$
is an independent, stationary, ergodic, centered, complex, Gaussian random process with power spectral density (PSD)
\begin{align}
\label{eq:GRP_flat_PSD}
S_\ell(f) = \begin{cases} 
\beta_\ell, & f \in [f_\ell-W_\ell,f_\ell+W_\ell], \\ 
0, & \text{else},
\end{cases}
\end{align}
where $W_\ell \in(0,\frac{1}{2})$ is the normalized half-bandwidth and $f_\ell \in [-\frac{1}{2},\ \frac{1}{2}]$ is the normalized center frequency. Here we assume that we observe the $x_\ell(t)$ on a set of uniform Nyquist rate samples such that the component signals of \eqref{eq:source_loc_cont} are given by $\vx_\ell[n] = x_\ell(n)$ for $n\in \{0,1,\dots,N-1\}$. By definition each $\vx_\ell \sim \mathcal{N}_\mathcal{C}(\mtx{0},\mR_\ell)$ where from \eqref{eq:GRP_flat_PSD} we have
\begin{align}
    \label{eq:GRP_cov_mat}
    \mR_\ell[m,n]  = \beta_\ell \int_{f_\ell-W_\ell}^{f_\ell+W_\ell}e^{j 2 \pi f (m-n)}df = 
    \begin{cases}
    \beta_\ell 2W_\ell, & m=n,\\
    \beta_\ell \frac{\sin(2\pi W_\ell (m-n))}{\pi(m-n)} e^{j 2 \pi f_\ell (m-n)}, & m\neq n.
    \end{cases}
\end{align}
If we let $\mE_{\ell}$ denote a diagonal matrix with entries $\mE_{\ell}[n,n] = e^{j 2 \pi f_\ell n}$ and 
\begin{align}
    \mB_{\ell} = \label{eq:prolate_matrix}
    \begin{cases}
    2W_\ell, & m=n,\\
    \frac{\sin(2\pi W_\ell (m-n))}{\pi(m-n)}, & m\neq n.
    \end{cases}
\end{align}
then $\mR_\ell = \beta_\ell \mE_{\ell}\mB_{\ell}\mE_{\ell}^H$. The motivation behind expressing the covariance matrix in this manner is that \eqref{eq:prolate_matrix} -- which is known in the literature as the \emph{prolate matrix} \cite{Varah93,Bojanczyk95} -- has many well-studied and favorable properties. The eigenvectors of $\mB_\ell$, denoted as $\vs_{\ell}^{(0)},\vs_{\ell}^{(1)},\dots,\vs_{\ell}^{(N-1)}$, are known as the \emph{Slepian basis} vectors. Their associated eigenvalues $\lambda_{\ell}^{(0)},\lambda_{\ell}^{(1)},\dots, \lambda_{\ell}^{(N-1)}$ are distinct and strictly between 0 and 1 with a particularly interesting clustering behavior \cite{SlepianV}: slightly fewer than $2NW_{\ell}$ eigenvalues are very close to 1 while slightly fewer than $N-2NW_{\ell}$ eigenvalues are very close to 0. The eigenvalues that do not fit into either of these clusters are provably few in number \cite{DPSSEig}. This means that the covariance matrix $\mR_\ell$ is well-approximated as being low-rank, and hence we can expect to be able to approximate $\vx_\ell$ using a low-dimensional subspace. 
% In particular, if $J=\ceil{2NW_{\ell}}$ the $k$th largest eigenvalue with $k\geq J$ satisfies\footnote{See\cite[Cor.\ 1]{DPSSEig} for a precise statement of \eqref{eq:eig_bound}.} 
% \begin{align}
%     \label{eq:eig_bound}
%     \lambda_{\ell}^{(k)} \leq c_1 \exp{\left( - \frac{k-J}{c_2}\right)}
% \end{align}
% where $c_1$ and $c_2$ are reasonably small known constants depending on $\log{(J+1)}$.

In particular, it is a well known result that the optimal MMSE orthobasis is given by the $K$-dominant eigenvectors of $\mR_\ell$\cite{Stark:1994}. Due to the clustering of the prolate matrix eigenvalues, choosing $K \approx 2NW_\ell$ produces a low-dimensional representation that captures all but a very small amount of the energy contained in $\vx_\ell$\cite{SlepianV,DPSSEig}. Letting $\mPsi_\ell = \mE_{\ell}\mS_{\ell,K}$ we will henceforth refer to $\mathcal{R}\{\mPsi_\ell\}$ as the $\ell^{\text{th}}$ Slepian space, and we can reasonably write 
\begin{align}
    \label{eq:slepian_rep}
    \vx_\ell \approx \mPsi_\ell\valpha_\ell
\end{align}
for a properly calculated vector of $K$ coefficients $\valpha_\ell$.

% To find such an approximation we consider the optimization program 
% \begin{align}
%     \label{eq:optimal_basis}
%     \minimize_{\mQ \in \mathbb{C}^{N\times K}} \mathbb{E} \norm{\vx_\ell - \mQ\mQ^H \vx_\ell}_2^2 \ \text{subject to} \ \mQ^H\mQ = \mtx{I}
% \end{align}
% which seeks to find the $K$-dimensional orthonormal basis that best represents $\vx_\ell$. It is a well known result that the solution to \eqref{eq:optimal_basis} is given by the $K$-dominate eigenvectors of $\mR_\ell$\cite{Stark:1994}. Letting $\mS_{\ell,K} = [\vs_{\ell}^{(0)} \ \vs_{\ell}^{(1)} \ \dots \ \vs_{\ell}^{(K-1)}]$ then $\mQ = \mE_{\ell}\mS_{\ell,K}$. From \eqref{eq:eig_bound} we can derive quantitative bound
% \begin{equation}
% 	\label{eq:taillambdak}
% 	\mathbb{E}\left[\|\vx_\ell - \mQ \mQ^H\vx_\ell\|_2^2\right] = \beta_\ell \sum_{k=K+1}^N\lambda_{\ell}^{(k)} \leq \beta_\ell c_3\exp\left(-\frac{K-J}{c_4}\right), \quad K\geq J,
% \end{equation}
% where $c_3$ and $c_4$ are dependent on $\log{(J+1)}$. What \eqref{eq:taillambdak} tells us is that the first $K\approx 2NW_\ell$ modulated Slepian basis vectors provides a low dimensional representation that captures all but a very small amount of the energy contained in $\vx_\ell$. Letting $\mPsi_\ell = \mE_{\ell}\mS_{\ell,K}$ we will henceforth refer to $\mathcal{R}\{\mPsi_\ell\}$ as the $\ell$th Slepian space, and we can reasonably write \begin{align}
%     \label{eq:slepian_rep}
%     \vx_\ell \approx \mPsi_\ell\valpha_\ell
% \end{align}
% for a properly calculated vector of $K$ coefficients $\valpha_\ell$.

We have established that each component $\vx_\ell$ admits an accurate low dimensional representation in a properly chosen Slepian space. To represent the superposition of signals in \eqref{eq:source_loc_cont} we utilize a union of Slepian spaces model, which requires the additional assumption that $|f_\ell-f_{\ell'}|>\max_{\ell} W_\ell+W_{\ell'}$ for all $\ell \neq \ell'$ such that the frequency bands do not overlap. Signals with these characteristics are often termed ``multi-band," and we will use this terminology for the remainder of the paper. Under this constraint, the PSDs of the individual processes simply superimpose such that
\begin{align*}
    S(f) = \sum_{\ell=1}^L S_\ell (f), ~ \mR = \sum_{\ell=1}^L \mR_\ell.
\end{align*}
Of course, the optimal low-dimensional representation in this case would be given by the first $K \approx 2N\sum_{\ell=1}^L W_\ell$ dominant eigenvectors of $\mR$, but synthesising such a basis requires full a priori knowledge of the statistics of $y(t)$. Noting that $\norm{\mPsi_{\ell}^H\mPsi_{\ell'}}$ is exceptionally small for $\ell \neq \ell'$ we are instead motivated to settle for a union of subspaces model \cite{davenport2012compressive}. The basic idea behind this is that since each $\vx_\ell$ admits a Slepian space representation it is reasonable to assume that
\begin{align}
    \label{eq:union_of_slep}
    \sum_{\ell=1}^L\vx_\ell \approx \sum_{\ell=1}^L\mPsi_{\ell} \valpha_\ell.
\end{align}
Letting $\valpha = [\valpha_1^T \ \valpha_2^T \ \dots \ \valpha_L^T]^T$ and $\mPsi = [\mPsi_{1} \ \mPsi_{2}\ \dots \ \mPsi_{L}]$ we can compactly represent the approximation in \eqref{eq:union_of_slep} as $\widehat\vx=\mPsi \valpha$. With this model in place, the broadband source localization problem amounts to determining the a priori unknown union of Slepian spaces that well represents samples of \eqref{eq:source_loc_cont}. This can be equivalently interpreted as determining the spectral support (i.e., the respective $(f_\ell,~W_\ell)$) of each $x_\ell(t)$.

One key respect in which the broadband source localization problem diverges from the traditional SLE framework is that we do not necessarily aim to accurately recover an underlying ``ground truth'' $\valpha_\ell$. In the narrowband SLE context, if we are able to accurately estimate the active frequencies, then estimating the corresponding $c_\ell$ in \eqref{eq:sinus_SLE} is relatively straightforward. % there are a small number of basis functions and coefficients that exactly represent the underlying signal in the narrowband case. 
This is not the case in the broadband setting when using Slepian space representations. In particular, depending on the number of Slepian basis vectors used in constructing our representations, the matrix $\mPsi$ may be ill-conditioned and hence the problem of estimating $\valpha$ from $\vy$ can be ill-posed, even when we are provided an exact estimate of the spectral support. In this context we do not need to worry about the precise estimate of $\valpha$ as long as  $\mPsi\valpha$ yields a good representation of the signal. %The same cannot be said for arbitrary broadband signals as even in a deterministic setting with the full temporal extent of the samples known simple Fourier analysis will yield a result that relies on a continuum of basis functions.\footnote{More explicitly the set $\{e^{j 2 \pi f n}~:~f\in[f_\ell-W_\ell,f_\ell+W_\ell]\}$ for each component signal.} Finite windowing of the samples merely relaxes this to requiring a set of $N$ basis functions for a perfect representation of the signal. There is no notion of a perfect low-dimensional representation of an arbitrary broadband signal  sampled over a finite window. Therefore we must always settle for an approximation and we do not particularly care what the $\valpha$ are so long as $\mPsi\valpha$ yields a good representation of the signal.

As a final note, our derivation of the union of Slepian space model followed from the assumption that each $x_\ell(t)$ is a Gaussian random process with flat PSD. However, this model generalizes well to ``typical" bandlimited signals. What is meant by ``typical" is that a collection of finite samples from said signals remain spectrally concentrated in-band. Though counter-examples can be constructed, signals that do not abide by this behavior are rarely seen in practice. As will be shown in the experimental results section, even when the signal is explicitly not a Gaussian random process it admits an accurate representation in $\mPsi$.

%%%%%%%%%%%%%%%%%%%%%%%%%%%%%%%%%%%%%%%%%%%%%%%%%%%%%%%%%%%
\subsection{Spectral leakage of broadband signals}
\label{sec:dictionary}

%Introduce Slepian space's and the role they play in forming an appropriate dictionary

A powerful benefit to the Slepian subspace model is that it is robust to ``spectral leakage." By spectral leakage, we mean the phenomenon where, due to truncation effects, the DFT of a finite window of a bandlimited signal will have significant sidelobes (nonzero DFT coefficients outside the signal's frequency band). A consequenece of this in a multiband signal is that sidebands of a component $\vx_\ell$ can obscure other component signals, particularly those with smaller relative power. Additionally, the DFT does not yield as low-dimensional of a representation for bandlimited and multiband signals. In this subsection, we elaborate on this latter point through a qualitative example that demonstrates the greater utility of the Slepian basis representation over a more standard DFT representation. %The former point will be discussed further in the next subsection. 

\begin{figure}[t]
\begin{center}
$$\vx_1[n] = 3\cos(\tfrac{2\pi \cdot 2}{100}n) - 2\cos(\tfrac{2\pi \cdot 5}{100}n) + 4\cos(\tfrac{2\pi \cdot 9}{100}n), \quad n = 0,1,\ldots,99$$
\begin{subfigure}[b]{0.49\textwidth}
\centering
\includegraphics[scale=0.30]{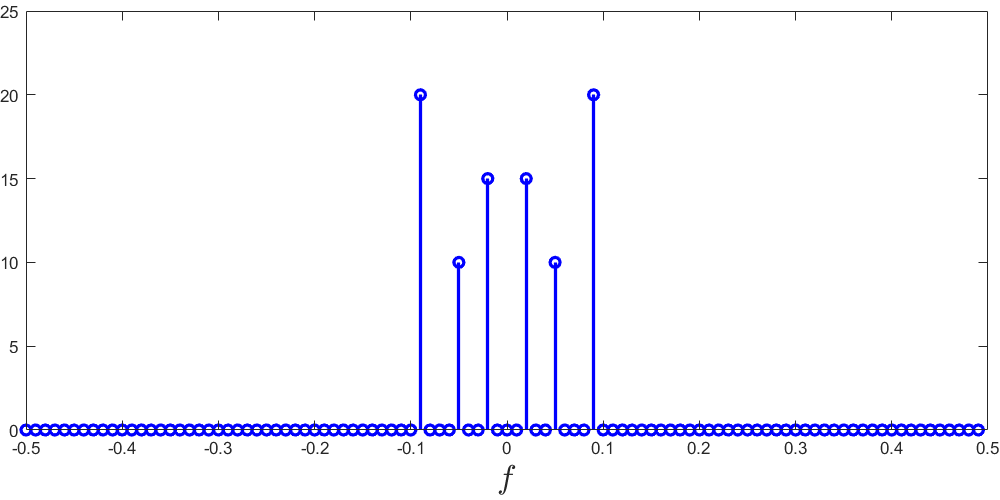}
\end{subfigure}
\begin{subfigure}[b]{0.49\textwidth}
\centering
\includegraphics[scale=0.30]{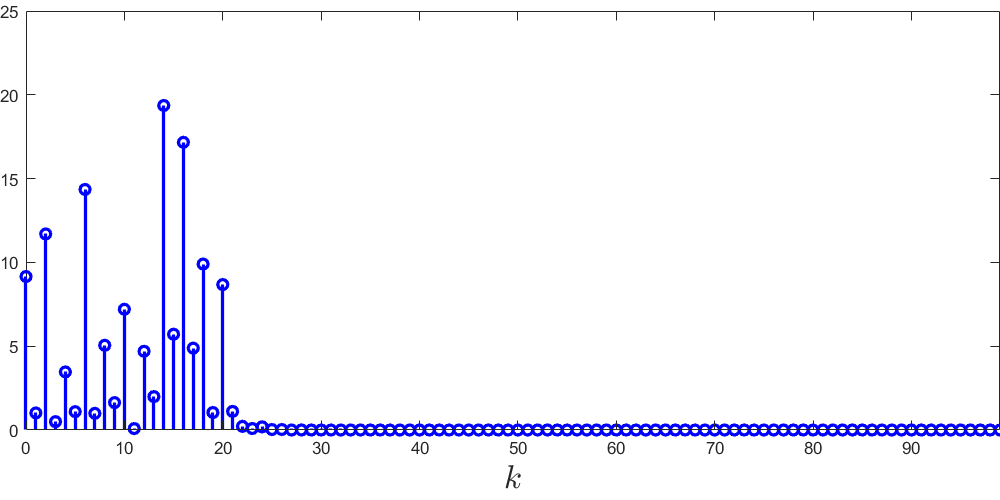}
\end{subfigure}

$$\vx_2[n] = 3\cos(\tfrac{2\pi \cdot 1.6}{100}n) - 2\cos(\tfrac{2\pi \cdot 5.1}{100}n) + 4\cos(\tfrac{2\pi \cdot 8.5}{100}n), \quad n = 0,1,\ldots,99$$
\begin{subfigure}[b]{0.49\textwidth}
\centering
\includegraphics[scale=0.30]{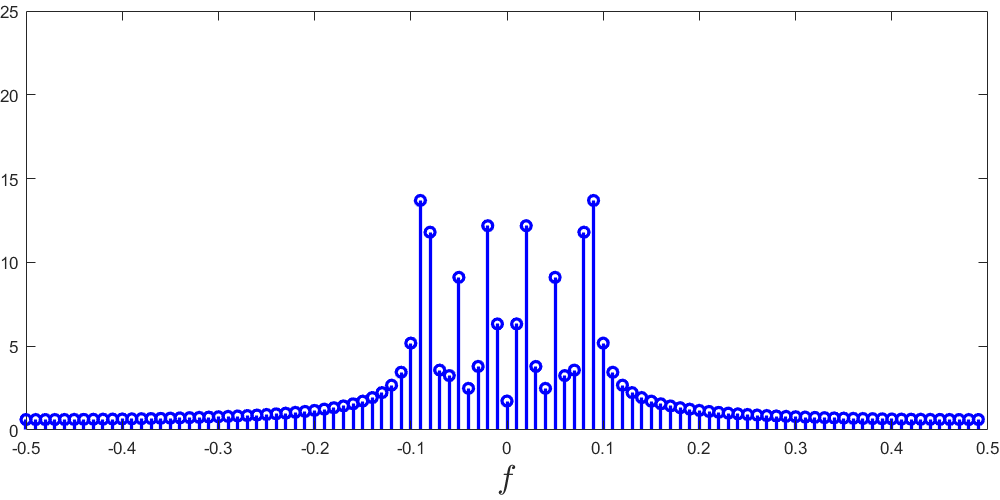}
\end{subfigure}
\begin{subfigure}[b]{0.49\textwidth}
\centering
\includegraphics[scale=0.30]{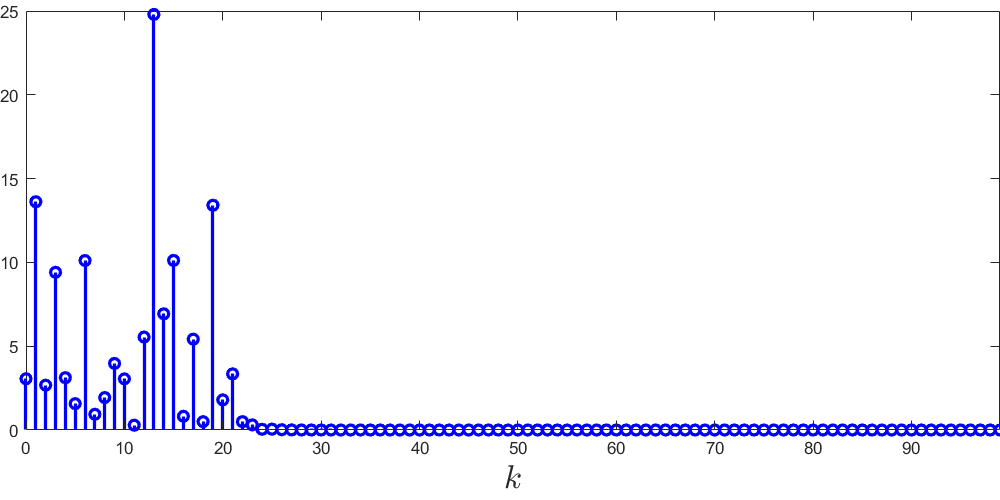}
\end{subfigure}
\caption{\small \sl(Top row) A plot of the magnitudes of the DFT coefficients (top left) and the Slepian basis coefficients with $N = 100$ and $W = \tfrac{1}{10}$ (top right) for the signal $\vx_1[n]$, which is a sum of three sinusoids at grid frequencies. (Bottom row) A plot of the magnitudes of the DFT coefficients (bottom left) and the Slepian basis coefficients with $N = 100$ and $W = \tfrac{1}{10}$ (bottom right) for the signal $\vx_2[n]$, which is a sum of three sinusoids at off-grid frequencies.}

\label{fig:Coefficients}
\end{center}
\end{figure}

Consider the top row of Figure~\ref{fig:Coefficients}, where we plot the magnitudes of the DFT coefficients and the magnitudes of the Slepian basis coefficients (with $N = 100$ and $W = 0.1$) of the discrete signal $\vx_1[n] = 3\cos(\tfrac{2\pi \cdot 2}{100}n) - 2\cos(\tfrac{2\pi \cdot 5}{100}n) + 4\cos(\tfrac{2\pi \cdot 9}{100}n)$ for $n = 0,1,\ldots,99$. In the bottom row of Figure~\ref{fig:Coefficients}, we plot the magnitudes of the DFT coefficients and the magnitudes of the Slepian basis coefficients (with $N = 100$ and $W = 0.1$) of the discrete signal $\vx_2[n] = 3\cos(\tfrac{2\pi \cdot 1.6}{100}n) - 2\cos(\tfrac{2\pi \cdot 5.1}{100}n) + 4\cos(\tfrac{2\pi \cdot 8.5}{100}n)$ for $n = 0,1,\ldots,99$. 

Since $\vx_1$ is a sum of three real sinusoids at grid frequencies, only $6$ DFT coefficients are non-zero. However, $\vx_2$ is a sum of three real sinusoids at off-grid frequencies. As a result, all $100$ DFT coefficients are non-zero. Furthermore, the largest $23$ and $67$ DFT coefficients capture $94.11\%$ and $99.02\%$ of the energy in $\vx_2$ respectively. In contrast, the first $23$ and $26$ Slepian basis coefficients capture $99.996\%$ and $99.99993\%$ of the energy in $\vx_1$ respectively. Also, the first $23$ and $26$ Slepian basis coefficients capture $99.993\%$ and $99.99997\%$ of the energy in $\vx_2$ respectively. Hence the Slepian basis does a significantly better job than the DFT basis at representing a discrete signal bandlimited to $|f| \le W$ even when the signal contains off-grid frequencies.

\subsection{Multi-tapered spectral estimation}
\label{sec:MTSE}

Above we have argued that a union of Slepian spaces model can be used to form highly accurate representations of multi-band signals. However, this begs the question of how to actually estimate the appropriate frequency bands so that we can build this model. The active frequency bands, modelled using the Slepian spaces spanned by $\mPsi_\ell$, can potentially lie on a continuum of possible frequencies $f_\ell$. If our signal consisted of a single active frequency band of known bandwidth, the ideal course of action seems clear: we would like to project $\vy$ onto all possible Slepian spaces (as parameterized by the center frequency of the band) to determine which one best represents the signal. Though this is a seemingly difficult task it can actually be achieved efficiently by applying multi-tapered spectral estimate (MTSE), often termed ``Thomson's multi-taper method" \cite{Thomson82}. In this subsection we overview Thomson's method and show how it can be used to estimate a signal's spectral support.

To perform MTSE we require choosing a bandwidth $W'$ to form a ``base" candidate subspace to test our signal against. Choosing $K'\approx 2NW'$ we form the set of candidate subspaces as $\mE_f \mS_{W',K'}$ for $f\in[0,1)$. We then determine the average energy contained in each of these subspaces by computing
\begin{align}
    \label{eq:MTSE}
    \widehat{S}_{\text{MTS}}(f) = \dfrac{1}{K'}\left\|\mE_{f}\mS_{W',K'}\mS_{W',K'}^H\mE_{f}^H\vx\right\|_2^2 = \dfrac{1}{K'}\sum_{k = 0}^{K'-1}\left|\sum_{n = 0}^{N-1}\vs_{W'}^{(k)}[n]\vx[n]e^{-j2\pi f n}\right|^2.
\end{align}
In essence,~\eqref{eq:MTSE} calculates the average energy $\vx$ contains in the frequency interval $[f-W',f+W']$. Therefore, by calculating $\widehat{S}_{\text{MTS}}(f)$ over a very fine grid of frequencies and observing where the peak (or peaks) reside we can estimate the locations of the active frequency band (or bands) and begin constructing an estimate of $\mPsi$. Of course the performance of the MTSE depends on our choice in $W'$ and subsequently $K'$.

\begin{figure}[t]
    \centering
    \includegraphics[width = 0.4\textwidth]{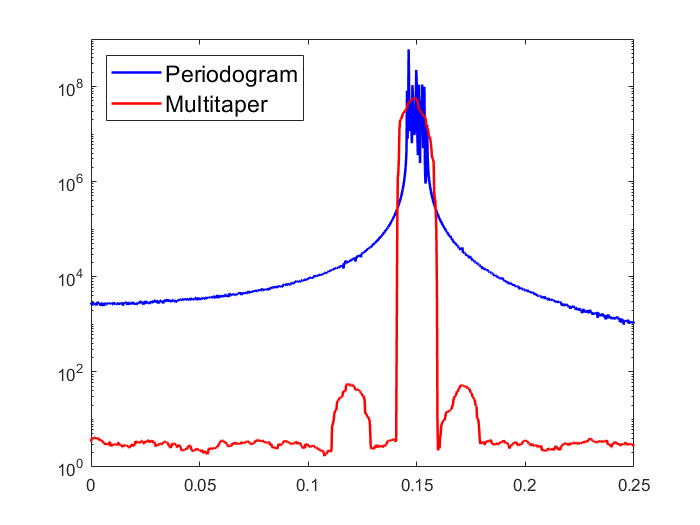}
    \caption{\small \sl A plot of the periodogram and the multitaper spectral estimate of a multiband signal with three components, one of which is significantly stronger than the other two.}
    \label{fig:MultitaperExample}
\end{figure}

In general we would like to choose $W'$ such that the spectral window generated by $\mE_{f}\mS_{W',K'}$ does not simultaneously capture energy from two distinct bands. Furthermore for $W_\ell>W'$ the range of $\mE_{f_\ell}\mS_{W',K'}$ is approximately contained in the range of $\mPsi_\ell$ with the reverse holding for $W'>W_\ell$. Hence $\widehat{S}_{\text{MTS}}(f_\ell)$ will still be large even when $W' \neq W_\ell$. Choosing the parameter $K'$ in the multiband setting is a somewhat context-dependent process that depends greatly on the dynamic range disparities between sources.\footnote{By dynamic range we mean how $\max_\ell \beta_\ell$ compares to $\min_\ell \beta_\ell$.} To guide this process, we note that the $\lambda_{W'}^{(k)}$ associated with each $\vs_{W'}^{(k)}$ tells us what fraction of the DTFT of $\vs_{W'}^{(k)}$ lives in the interval $[-W',W']$~\cite{SlepianV,FastMultitaper,FST}. When $\lambda_{W'}^{(k)}\approx 1$ almost all of the energy is contained in $[-W',W']$, while when $\lambda_{W'}^{(k)}\approx 0$ almost all of the energy resides outside the interval. Therefore when the dynamic range disparity is drastic we may choose $K'$ to be slightly less that $2NW'$ such that \eqref{eq:MTSE} only measures the energy predominately in the range $[f-W',f+W']$ in order to mitigate any impact of sidelobes from the larger source. As an example, in Figure~\ref{fig:MultitaperExample} we generate a (noiseless) vector $\vy$ with samples of a multiband signal with three components, one of which is several orders of magnitude stronger than the other two. We plot both the periodogram (squared magnitude of the DFT coefficients) of $\vy$ and the multitaper spectral estimate $\widehat{S}(f_c)$ for $K'$ slightly less than $2NW'$. We note that the spectral leakage phenomenon causes the strongest source to have sidelobes which obscure two weaker sources. However, all three sources are clearly visible from the multitaper estimate. 

Finally, we note that there are computationally efficient methods for working with the Slepian basis vectors when the signal length $N$ is large. In \cite{Gruenbacher94}, a method is developed for efficiently computing Slepian basis vectors by exploiting the fact that the prolate matrix commutes with a tridiagonal matrix \cite{SlepianV}. More recently, in~\cite{FST}, it is shown that computing the projection of a vector $\vy$ onto the span of the first $\approx 2NW$ Slepian basis vectors can be done in $O(N \log N \log \tfrac{1}{\eps})$ operations via an FFT plus a low-rank correction, which is a significant improvement over the $O(WN^2)$ operations necessary to compute the projection naively. Leveraging this approach, in \cite{FastMultitaper}, it is shown that the multitaper spectral estimate of $\vy$ can be evaluated at a grid of $N$ frequencies in $O(N \log^2 N \log \tfrac{1}{\eps})$ operations, as opposed to the $O(WN^2\log N)$ operations necessary to compute the multitaper spectral estimate directly. By exploiting these fast methods, we can greatly decrease the time and memory requirements of our broadband source localization algorithms.
%%%%%%%%%%%%%%%%%%%%%%%%%%%%%%%%%%%%%%%%%%%%%%%%%%%%%%%%%%%
\subsection{The necessity of adjusting for bandwidth}

Given the somewhat more complex nature of our approach to broadband signal estimation, one may naturally wonder when it is truly necessary, and when the narrowband assumption used in classical SLE methods will be sufficient. Of course, in practice the ideal narrowband assumption of \eqref{eq:sinus_SLE} (i.e., pure tones) is never true, but one might expect that it is often a reasonable approximation. Indeed, when the bandwidth of the sources is not appreciable, applying the standard SLE algorithms will still yield reasonable results. This begs the question ``How broadband is too broadband?" We offer a simple experiment to establish when the bandwidth is considered appreciable, necessitating a broadband model. 

\begin{figure}[t]
    \centering
    \includegraphics[width=.5\textwidth]{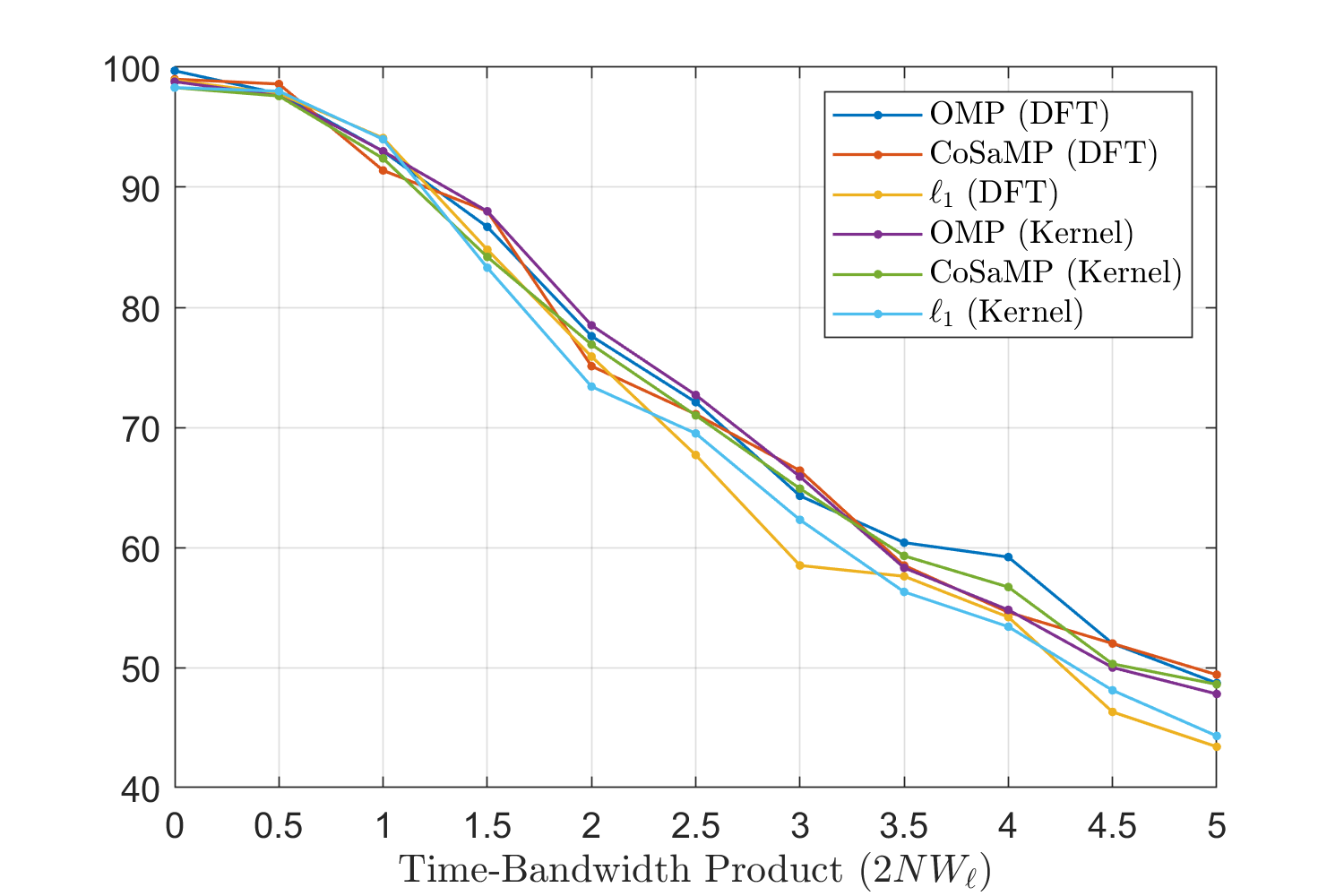}
    \caption{\small \sl Average percent of sources detected for a for a variety of SLE methods as a function of the individual sources' time-bandwidth product. The parenthesis indicate whether a discrete Fourier (DFT) or smoothed kernel (Kernel) basis was utilized in the respective algorithm. As the time-bandiwdth product increases we see a similar degradation in performance across all selected methods.}
    \label{fig:SLE_tbp_comparison}
\end{figure}

We set $N=2^{10}$ and generate $L=5$ noiseless broadband sources such that we are operating in a regime where we are confident narrowband SLE algorithms work. We examine two versions of the CoSaMP, OMP, and $\ell_1$-minimization algorithms, where the first set uses standard DFT basis representations while the second utilizes a ``smoothed" DFT basis. The latter case forms its basis by modulating some bandlimited kernel as described in \cite{Eftekhari:2015}. For our experiments we choose this kernel to be the first Slepian basis vector generated from a prolate matrix with a time-bandwidth product of 1. The results presented in Figure \ref{fig:SLE_tbp_comparison} demonstrates that as $2NW_\ell$ exceeds 1 for each source there is a roll off in performance. This leads us to conclude that for signals in which the time bandwidth product exceed 1 the bandwidth has become appreciable to the point where the narrowband assumption simply fails. As further justification for why this is the case we note that the time-bandwidth product acts as an estimate of the signal's underlying degrees of freedom\cite{SlepianV}. Since the SLE algorithms account for only a single degree of freedom in the support of each source it is not surprising that this performance roll off occurs. This justifies our motivation for developing a purely broadband centric method of source localization.

%%%%%%%%%%%%%%%%%%%%%%%%%%%%%%%%%%%%%%%%%%%%%%%%%%%%%%%%%%%
\section{Broadband Signal Estimation}

\subsection{Preliminaries}

While the MTSE described in Section~\ref{sec:MTSE} can be an effective approach to estimating the spectral support of a broadband signal, one of the key messages of the recent literature on SLE summarized in Section~\ref{sec:spectral_line_est} is that simple spectral estimation techniques can often be improved upon by leveraging techniques from the sparse approximation literature. In particular, iterative algorithms such as OMP and CoSaMP can often improve on simple ``peak picking'' approaches, particularly when the active bands are close to each other or exhibit high dynamic range so that weaker components may be ``hidden'' in the side lobes of larger ones.   Therefore, our proposed broadband source localization algorithms are largely modified versions of OMP and CoSaMP. Similar extensions for any of the other popular iterative sparse approximation algorithms would be possible. 

To allow for a more compact summary of these algorithms in subsequent sections, we first define a series of useful functions:
\begin{itemize}
    \item \underline{$P_{\Delta f}(\mtx{S}_f,K)$}: Takes a sampled spectral estimate $\mtx{S}_f$ and returns a vector $\vf$ containing the normalized center frequency estimates corresponding to the $K$ largest peaks. A minimum spacing of $\Delta f$ between component center frequencies is assumed such that no two estimates are within $\Delta f$ of each other.
    \item \underline{$G_{\Delta f}(\mtx{S}_f,\mtx{f})$}: Takes a sampled spectral estimate $\mtx{S}_f$ and a vector of estimated center frequencies $\mtx{f}$ and returns a vector of estimated bandwidths $\mtx{W}$ and a vector of updated center frequencies that have been adjusted to account for the band edges. Once again, we utilize the minimum separation parameter $\Delta f$ in order to limit the search region in the bandwidth estimation step. Without this parameter, what can be considered as a single band/component becomes ambiguous.
    \item \underline{$D_{\epsilon}(\mtx{f},\mtx{W})$}: Takes a set of center frequencies and bandwidths and returns a $\mPsi$ that spans these regions of the spectrum. Some of the proposed methods of $\mPsi$ formation require an approximation parameter $\epsilon$ in order to specify the degree of precision. 
    \item \underline{$\mathcal{S}_{\mtx{\Psi}}(\mtx{z},K)$ }: Takes an estimate $\mtx{z}$ of the signal in the range of $\mPsi$ and returns the center frequencies and bandwidths of the most ``significant" sources. This encompasses everything associated with the ``pruning'' step that occurs at various stages in either algorithm. This can be done in several ways, but we utilize the approach in Algorithm~\ref{alg:pruning}.
\end{itemize}
As will be demonstrated in Section~\ref{sec:simulation_results} these algorithms can be applied in the context where instead of observing $\vy$ directly we instead observe \emph{compressed} observations. To explicitly incorporate this into the sampling model we incorporate an $M\times N$ sensing matrix $\mA$ such that
\begin{align}
    \label{eq:source_loc_comp}
    \vy = \mA\sum_{\ell=1}^L\vx_\ell+\veta.
\end{align}
Of course when $\mA=\mtx{I}$, the model in~\eqref{eq:source_loc_comp} and directly sampling as in~\eqref{eq:source_loc_cont} are equivalent, but below we include $\mA$ to allow for full generality.

\begin{algorithm}[t]
\caption{$\mathcal{S}_{\mtx{\Psi}}(\mtx{z},K)$}
\label{alg:pruning}
\begin{algorithmic}[1]
\State {Input:}~$\vz,K$
\State $S(f) \gets \frac{1}{K} \norm{\mtx{S}_{K',W'}^{H}\mtx{E}_f^H \vz}_2^2$ \Comment{Multi-taper spectral estimate}
\State $\mtx{S}_f[n] \gets S(\frac{2 \pi n}{Nd})$ \Comment{Sample over a fine grid}
\State $\widehat{\mtx{f}} \gets P_{\Delta f}(\mtx{S}_f,K)$ \Comment{Identify $K$ largest peaks}
\State $\mtx{f} ,\,\mtx{W}  \gets G_{\Delta f}(\mtx{S}_f,\, \widehat{\vf})$ \Comment{Estimate bandwidth/Refine center frequencies}
\end{algorithmic}
\end{algorithm}

\subsection{Broadband OMP}

The broadband OMP algorithm follows the same general approach as the SLE implementation with a few additional features. A key step in the algorithm is the projection of the signal onto the model family via a MTSE. Other important deviations from the standard is use of a bandwidth estimation function that is coupled with an adaptive dictionary generation step. Pseudo-code for the algorithm is presented in Algorithm \ref{alg:CoSaMP_compressed}. The various hyperparameters that have not been mentioned thus far are discussed thoroughly below in Section~\ref{sec:hyper}.

\begin{algorithm}[t]
\caption{OMP for Broadband Signals}
\label{alg:OMP_compressed}
\begin{algorithmic}[1]
\State {Input:~Data} $\mtx{y}$,~{Number of sources/components} $L$
\State {Initialize:} $\mtx{r}^{(0)} \gets \mtx{y}$, $\mtx{\Psi}^{(0)} \gets \emptyset$, $i \gets 0$
\While{Not Converged}
\State $i \gets i+1$
\State $\mtx{v} = \mtx{A}^H\mtx{r}^{(i-1)}$ \Comment{Form proxy}
\State $S(f) \gets \frac{1}{K'} \norm{\mtx{S}_{K',W'}^{H}\mtx{E}_f^H \mtx{v}}_2^2$ \Comment{Multi-tapered spectral estimate}
\State $\mtx{S}_f[n] \gets S(\frac{2 \pi n}{Nd})$ 
\State $\widehat{\mtx{f}}_{1} \gets P_{\Delta f}(\mtx{S}_f,1)$ \Comment{Identify largest peak}
\State $\mtx{f}_{1} ,\,\mtx{W}_{1}  \gets G_{\Delta f}(\mtx{S}_f,\, \widehat{\mtx{f}}_{1})$ \Comment{Estimate bandwidth/Refine center frequency}
\State $\mtx{\Psi}_{1} \gets D_{\epsilon}(\mtx{f}_{1},\mtx{W}_{1})$
\State $\mtx{\Psi}^{(i)} \gets [\mtx{\Psi}^{(i-1)}\, \mtx{\Psi}_{1}]$ \Comment{Merge support}
\State $\mtx{f}^{(i)} \gets [\mtx{f}^{(i-1)}\, \mtx{f}_{1}]$ 
\State $\mtx{W}^{(i)} \gets [\mtx{W}^{(i-1)}\, \mtx{W}_{1}]$ 
\State $\mtx{x}^{(i)} \gets \mtx{\Psi}^{(i)}(\mtx{\Psi}^{(i)H}\mtx{A}^H\mtx{A}\mtx{\Psi}^{(i)}+\gamma \mtx{I})^{-1}\mtx{\Psi}^{(i)H}\mtx{A}^H\mtx{y}$ \Comment{Estimation}
\State $\mtx{r}^{(i)} \gets \mtx{y} - \mtx{A}\mtx{x}^{(i)} $ \Comment{Update residual}
\EndWhile \label{euclidendwhile}
\State $\mtx{f} ,\,\mtx{W} \gets \mathcal{S}_{\mtx{\Psi}^{(i)}}(\widehat{\mtx{x}}^{(i)},L)$ \Comment{Pruning}
\end{algorithmic}
\end{algorithm}

\subsection{Broadband CoSaMP}

The CoSaMP version of the algorithm is somewhat more intricate in that it picks multiple peaks at a time and then ``prunes" them prior to updating the residual. The inclusion of a signal space estimation and pruning step follows a similar methodology to \cite{Davenport:SSCoSaMP}. Otherwise the algorithm has modifications analogous to those made for OMP. Pseudo-code is presented in Algorithm \ref{alg:CoSaMP_compressed}.

\begin{algorithm}[t]
\caption{CoSaMP for Broadband Signals}
\label{alg:CoSaMP_compressed}
\begin{algorithmic}[1]
\State {Input:~Data} $\mtx{y}$,~{Number of sources/components} $L$
\State {Initialize:} $\mtx{r}^{(0)} \gets \mtx{y}$, $\mtx{\Psi}^{(0)} \gets \emptyset$, $i \gets 0$
\While{Not Converged}
\State $i \gets i+1$
\State $\mtx{v} = \mtx{A}^H\mtx{r}^{(i-1)}$ \Comment{Form proxy}
\State $S(f) \gets \frac{1}{K'} \norm{\mtx{S}_{K',W'}^{H}\mtx{E}_f^H \mtx{v}}_2^2$ \Comment{Multi-tapered spectral estimate}
\State $\mtx{S}_f[n] \gets S(\frac{2 \pi n}{Nd})$ 
\State $\widehat{\mtx{f}}_{2L} \gets P_{\Delta f}(\mtx{S}_f,2L)$ \Comment{Identify $2L$ largest peaks}
\State $\mtx{f}_{2L} ,\,\mtx{W}_{2L}  \gets G_{\Delta f}(\mtx{S}_f,\, \widehat{\mtx{f}}_{2L})$ \Comment{Estimate bandwidth/Refine center frequencies}
\State $\mtx{\Psi}_{2L} \gets D_{\epsilon}(\mtx{f}_{2L},\mtx{W}_{2L})$
\State $\widehat{\mtx{\Psi}} \gets [\mtx{\Psi}^{(i-1)}\, \mtx{\Psi}_{2L}]$ \Comment{Merge support}
\State $\widehat{\mtx{f}} \gets [\mtx{f}^{(i-1)}\, \mtx{f}_{2L}]$ 
\State $\widehat{\mtx{W}} \gets [\mtx{W}^{(i-1)}\, \mtx{W}_{2L}]$ 
\State $\widehat{\mtx{x}} \gets \widehat{\mtx{\Psi}}(\widehat{\mtx{\Psi}}^H\mtx{A}^H\mtx{A}\widehat{\mtx{\Psi}}+\gamma \mtx{I})^{-1}\widehat{\mtx{\Psi}}^H\mtx{A}^H\mtx{y}$ \Comment{Estimation}
\State $\mtx{f}^{(i)} ,\,\mtx{W}^{(i)}  \gets \mathcal{S}_{\widehat{\mtx{\Psi}}}(\widehat{\mtx{x}},L)$ \Comment{Pruning}
\State $\mtx{\Psi}^{(i)} \gets D_{\epsilon}(\mtx{f}^{(i)} ,\,\mtx{W}^{(i)})$
\State $\mtx{r}^{(i)} \gets \mtx{y} - \mtx{A}\mtx{\Psi}^{(i)}(\mtx{\Psi}^{(i)H}\mtx{A}^H\mtx{A}\mtx{\Psi}^{(i)}+\gamma \mtx{I})^{-1}\mtx{\Psi}^{(i)H}\mtx{A}^H\mtx{y}$ \Comment{Update residual}
\EndWhile \label{euclidendwhile}
\end{algorithmic}
\end{algorithm}
%\section{Practical Considerations}
%\label{sec:practical_considerations}
\subsection{Bandwidth estimation}
Up to this point we have ignored the details of the the bandwidth estimation algorithm $G_{\Delta f}(\mtx{S}_f,\mtx{f})$. Here we will discuss our proposed maximum likelihood based method of bandwidth estimation. We begin by considering the fully sampled version of \eqref{eq:source_loc_comp} in which $\mA=\mtx{I}$ and $\veta \sim \mathcal{N}_\mathcal{C}(\mtx{0},\sigma^2 \mtx{I})$. The associated sampled spectral estimate is denoted by $\mtx{S}_f$. In this scenario each sample of $\mtx{S}_f$ can be viewed as being generated according to one of two hypotheses: 
\begin{align*}
\mathcal{H}_0:& \text{~No signal present at this frequency,} \\
\mathcal{H}_1:& \text{~Signal present at this frequency.}
\end{align*}
We define the distribution under either hypothesis as $S_i \,|\, \mathcal{H}_0 \sim g(\mtx{S}_f[i] \, |  \, \mathcal{H}_0,\,\mu_0,\,\sigma_0)$ and $S_i \,|\, \mathcal{H}_1 \sim g(\mtx{S}_f[i] \, |  \, \mathcal{H}_1,\allowbreak\,\mu_1,\,\sigma_1)$ where $S_i$ is the random variable associated with a sample $\mtx{S}_f[i]$ and $g$ is a probability density function parameterized by $\mathcal{H}$, $\mu$, and $\sigma$. Our goal is to identify intervals of the samples in $\mS_f$ as being generated according to one of these hypothesis. %The distinct distributions under either hypothesis motivates to explore a modified version of the MLE.

To compute the likelihood that a particular interval of samples were drawn according to either hypothesis, we modify the standard log-likelihood definitions to include two integer index arguments $N_1$ and $N_2$ such that
\begin{align*}
    \mathcal{L}_0(N_1,N_2,\mu,\sigma | \mS_f) &= \sum_{i=N_1}^{N_2} \log{g(\mS_f[i] \, | \, \mathcal{H}_0,\,\mu,\,\sigma)},  \\
    \mathcal{L}_1(N_1,N_2,\mu,\sigma| \mS_f) &= \sum_{i=N_1}^{N_2} \log{g(\mS_f[i] \, | \, \mathcal{H}_1,\,\mu,\,\sigma)}.
\end{align*}
In general $\mu$ and $\sigma$ are unknown a priori, but we can estimate them using the sample mean and variance over the given interval which we denote by $\widehat{\mu}_{N_1,N_2}$ and $\widehat{\sigma}_{N_1,N_2}$ respectively. Let $\mtx{w}$ be a window of $N_w$ samples from $\mS_f$ centered at $\widehat{f}_{c}$. Our detection scheme assumes samples in $\mtx{w}$ under $\mathcal{H}_1$ lie in an interval of frequency bin indices $[n_1, n_2]$ while samples under $\mathcal{H}_0$ exist on the indices $\{[1, \, n_1) \cup (n_2, \,N_w]\}$, we then seek to find this partition, i.e., to estimate $n_1$ and $n_2$. Given two frequency indices $n_1$ and $n_2$ we define the composite log-likelihood as
{\small
\begin{align*}
    \widetilde{\mathcal{L}}_{\mu,\sigma}(n_1,n_2|\mtx{w}) = \mathcal{L}_0(1,n_1,\widehat{\mu}_{1,n_1},\widehat{\sigma}_{1,n_1}|\mtx{w})+ \mathcal{L}_1(n_1,n_2,\widehat{\mu}_{n_1,n_2},\widehat{\sigma}_{n_1,n_2}|\mtx{w})+  \mathcal{L}_0(n_2,N_w,\widehat{\mu}_{n_2,N_w},\widehat{\sigma}_{n_2,N_w}|\mtx{w}).
\end{align*}}
The bandwidth is then determined by computing the maximum likelihood estimate, i.e., solving the optimization program 
\begin{align*}
    \maximize_{n_1,n_2} ~\widetilde{\mathcal{L}}_{\mu,\sigma}(n_1,n_2|\mtx{w}).
\end{align*}

\begin{figure}[t]
\centering
\captionsetup[subfigure]{labelformat=empty}
\begin{subfigure}{.4\textwidth}
\centering
\includegraphics[width=\textwidth]{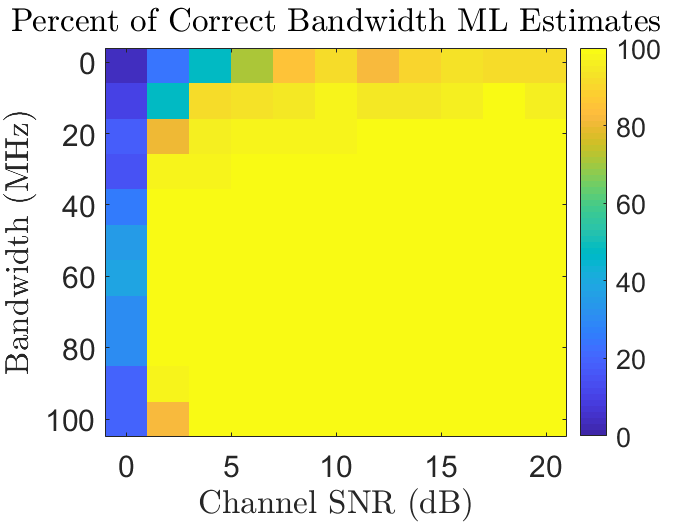}
\caption{(a)}
\end{subfigure}%
\hfil
\begin{subfigure}{.4\textwidth}
\centering
\includegraphics[width=\textwidth]{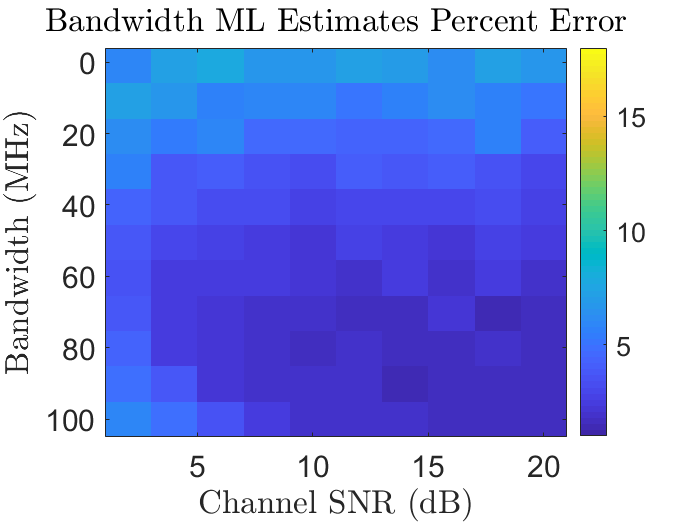}
\caption{(b)}
\end{subfigure}%
\caption{\small \sl Results of bandwidth estimation numerical experiments depicting (a) percent of correctly detected bandwidths (b) average percent error in correctly detected bandwidths. The vertical axis represents the true bandwidth of the generated signal and the horizontal axis represents the corresponding channel (in band) SNR.}
\label{fig:BW_Est_res}
\end{figure}

Despite its simplicity, this method is fairly robust. As a demonstration, we generated a series of bandlimited signals via a sum of sinusoids model\footnote{This model will be elaborated upon in Section~\ref{sec:simulation_results}, but can also be found in \cite{Tropp:2010}.} with $N=100 \, 000$. Gaussian noise was then added to the signal and a periodogram was used as the spectral estimate such that the samples are Rayleigh distributed and parameterized by some $\sigma$. The results of these experiments are presented in Figure \ref{fig:BW_Est_res} where the SNR and bandwidth of the test signals were varied. Each SNR-bandwidth pair was averaged over 100 test iterations with a successful detection defined as an estimate with less than $20 \% $ error. As is apparent from Figure \ref{fig:BW_Est_res}(a), for the vast majority of scenarios in which the SNR was greater than 0 our apporach leads to near perfect detection results. Of those correctly detected sources, Figure \ref{fig:BW_Est_res}(b) shows that percent error is low.

\subsection{Choosing hyperparameters}
\label{sec:hyper}

There are a variety of hyperparameters that have to be chosen in order for the algorithms above to behave as desired. Here we discuss some of the more important parameters and how they factor into performance.
\begin{itemize}
    \item \underline{$W'$}: This should be set to some lower bound on the half-bandwidth of the components of the multiband signal. Increasing $W'$ will bias the bandwidth and center frequency estimate in a manner that is difficult to refine due to the trapezoidal features.
    \item \underline{$K'$}:  We choose a cutoff parameter $\epsilon_c\in(0,1)$ such that $\lambda_{k'} > 1 - \epsilon_{\text{c}}$. Since $\lambda_{k'}$ is the fraction of in-band energy, in the case of high dynamic range $\epsilon_{\text{c}}$ should be small to maximize spectral isolation. However, a larger $\epsilon_{\text{c}}$ will result in a greater coherent SNR gain.
    \item \underline{$\gamma$}: This is the regularization parameter used in the least squares step of the OMP and CoSaMP algorithm. Depending on a variety of factors, namely the proximity of sources, $\mtx{\Psi}$ may be very poorly conditioned. The addition of this regularization parameter makes the least squares problem more stable at the cost of biasing the answer. This should be proportional to the SNR.
    \item \underline{$\mathcal{S}_{\mtx{\Psi}}(\mtx{z},K)$}: For our experiments we use Algorithm~\ref{alg:pruning}. However, there are a variety of metrics that could be used to compare different components. For certain scenarios it may be more prudent to implement an alternative ad hoc style of pruning.
\end{itemize}
There are a handful of other hyperparameters in the implementation, but the above are the ones that most contribute to the variability of performance. Parameters are generally compatible with both CoSaMP and OMP, meaning if one algorithm works well with a particular set of hyperparameters, so will the other with minimal re-tuning. In other words, these parameters are relatively algorithm independent, depending primarily on the details of the underlying source localization problem.
\section{Numerical Experiments: Spectral Line Estimation}
\label{sec:numerical_SLE}

Before beginning a thorough evaluation of our proposed algorithms, we first compare the performance of some selected methods for SLE in terms of their ability to estimate the spectral support of the component signals (i.e., how closely they approximate each $f_\ell$). For our experiments, we use $N=1024$ uniform samples of a sum of $L=5$ sinusoidal sources sampled at $f_s = 25$~GHz. We ensure that the sources are separated by at least $488$~MHz (20 grid points). The frequency and complex amplitude of the signals are drawn randomly. In our experiments, we simulate 100 realizations under these parameters and evaluate the algorithms based on the percentage of sources correctly detected as well as the relative $f_\ell$ estimation error. A source is considered to be correctly detected if the identified frequency is within $72$~MHz (3 grid points), and only correctly detected sources are included in the $f_\ell$ estimation error calculation. For algorithms that utilize an oversampled DFT dictionary (i.e., CoSaMP, OMP, and $\ell_1$-minimization) we choose the dictionary to be $10\times$ oversampled. We study the performance of the algorithms in noisy regimes where $\veta \sim \mathcal{N}_{\mathcal{C}}(\mtx{0},\sigma^2\mtx{I})$\footnote{$ \mathcal{N}_{\mathcal{C}}$ denotes a complex Gaussian distribution} by varying the signal-to-noise ratio (SNR).

\begin{figure}[t]
    \centering
    \includegraphics[width=.4\textwidth]{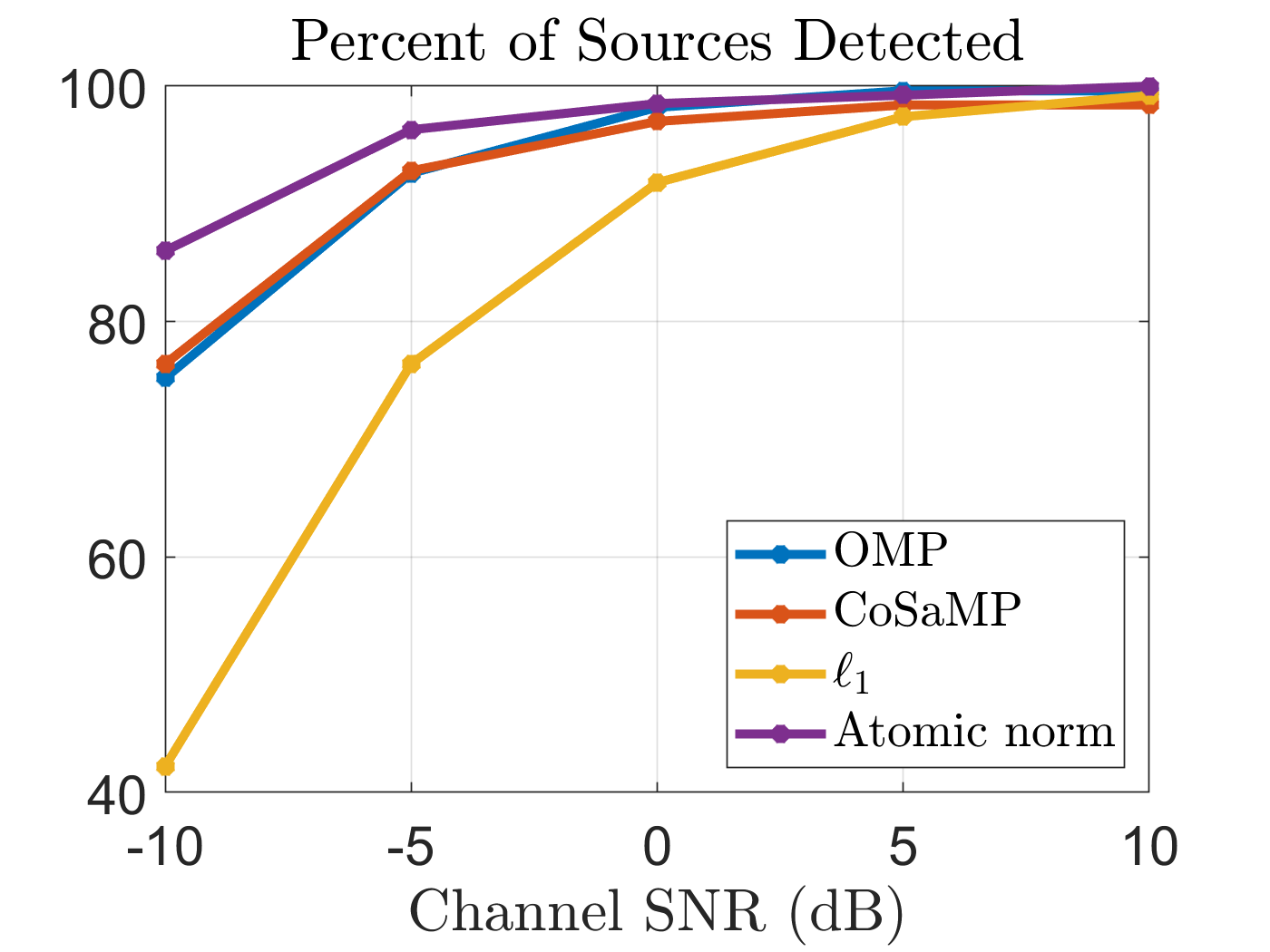}\hfil
    \includegraphics[width=.4\textwidth]{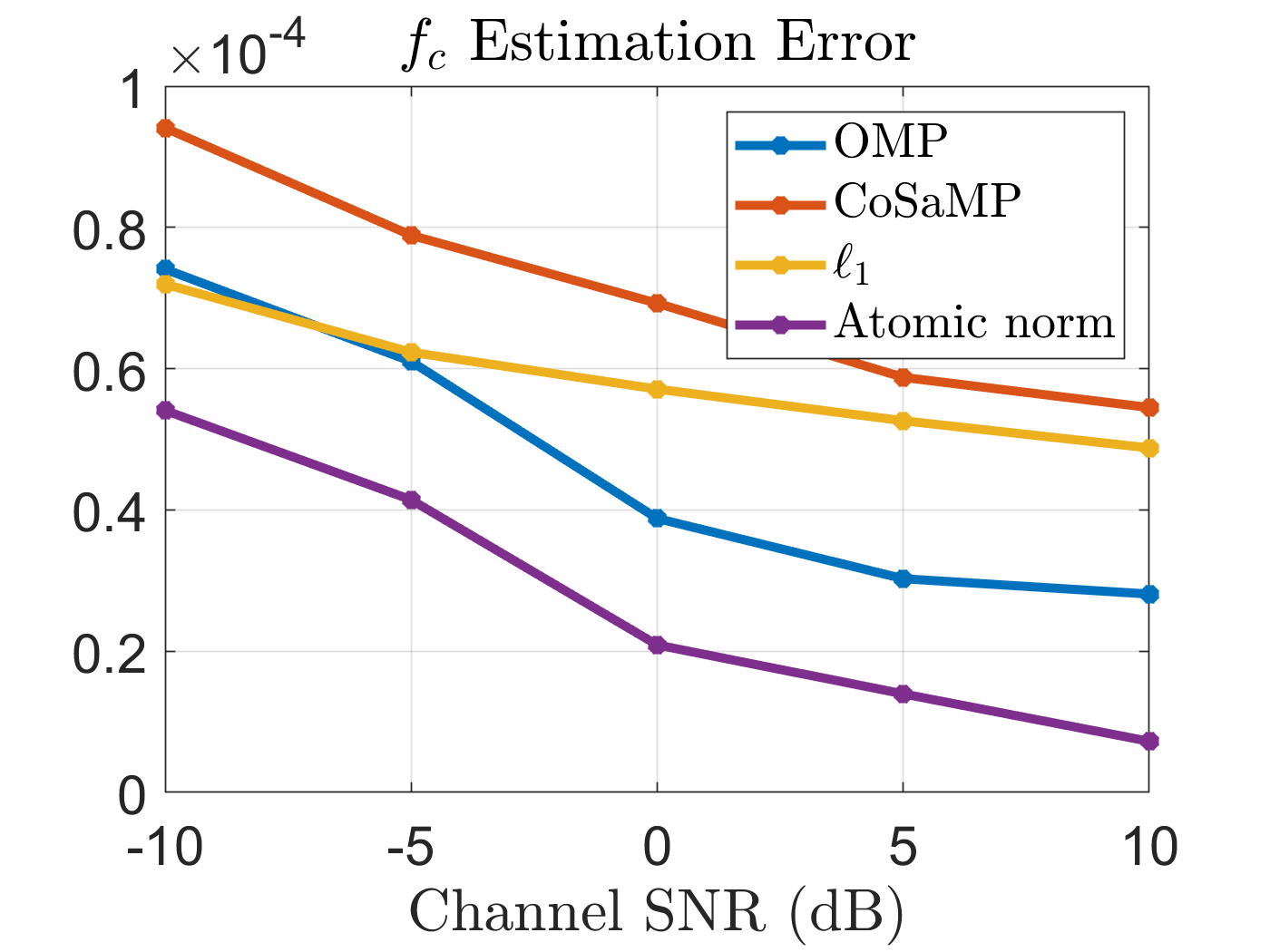}\hfil
    \caption{\small \sl  Comparison of SLE performance of OMP, CoSaMP, $\ell_1$-minimization, and atomic norm denoising algorithms under a variety of SNRs for a $N=1024$ signal. (left) Shows the percent of sources detected while (right) displays center frequency estimation error normalized by $f_s$. We note that all data points are averaged over 100 trials with exception to atomic norm denoising, which is averaged over 30 trials.}
    \label{fig:SLE_iterative_comparison}
\end{figure}

The results for OMP, CoSaMP, $\ell_1$-minimzation, and atomic norm minimization are shown in Figure~\ref{fig:SLE_iterative_comparison}. Due to the immense computational requirement of atomic norm minimization, we average over only 30 trials in that case. Nevertheless the atomic norm minimization framework demonstrates a clear trend that establishes it as the most reliable and accurate method of detection. However, the poor computational scaling of this method is enough to make it impractical even for a relatively short signal length. We also see that CoSaMP and OMP perform at comparable levels and generally outperform $\ell_1$-minimization. 

\begin{figure}[th]
    \centering
    \begin{subfigure}[b]{.4\textwidth}
    \centering
    \includegraphics[width=\textwidth]{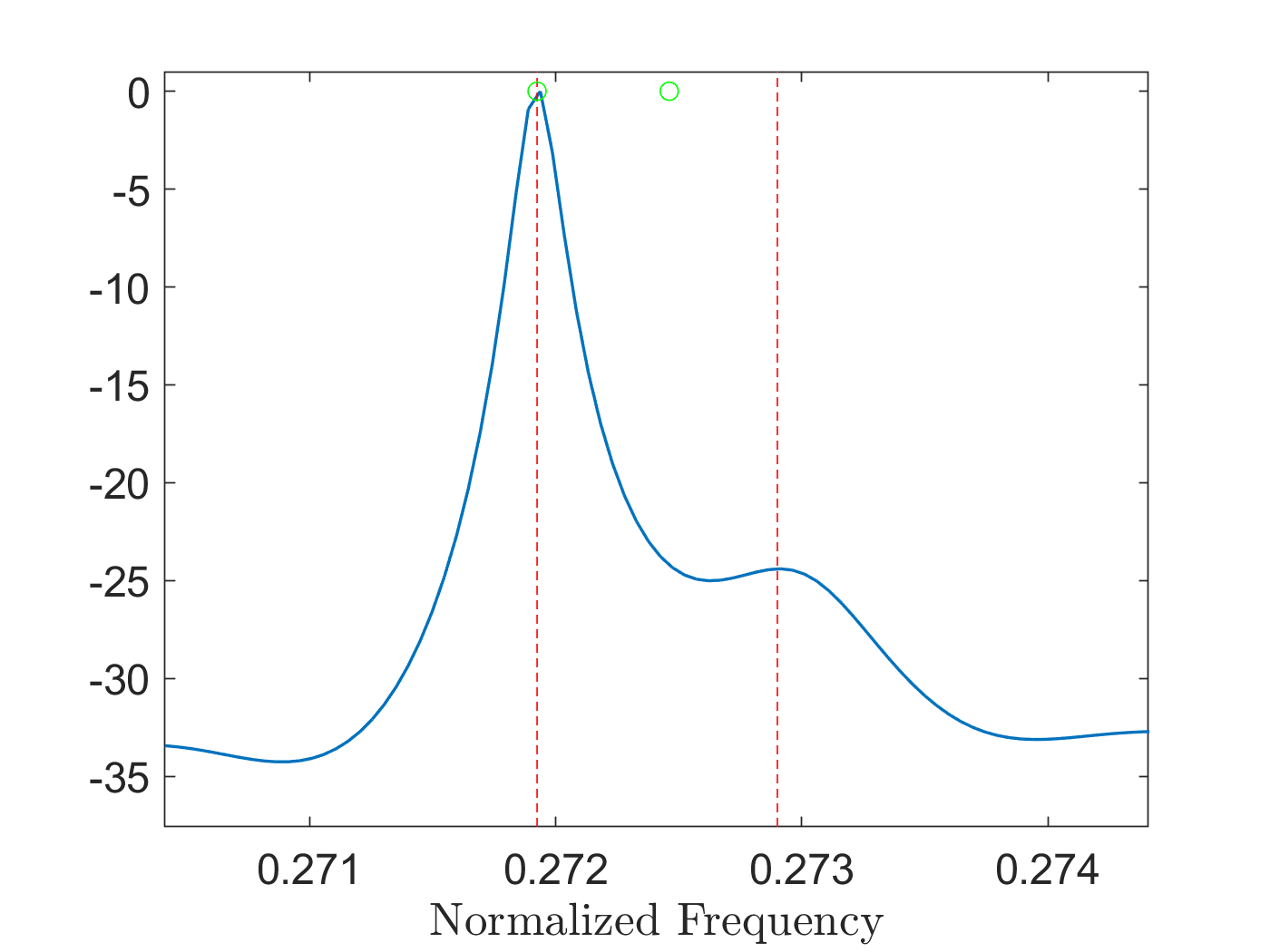}
    \caption{}
    \end{subfigure}
    \hfil
    \begin{subfigure}[b]{.4\textwidth}
    \centering
    \includegraphics[width=\textwidth]{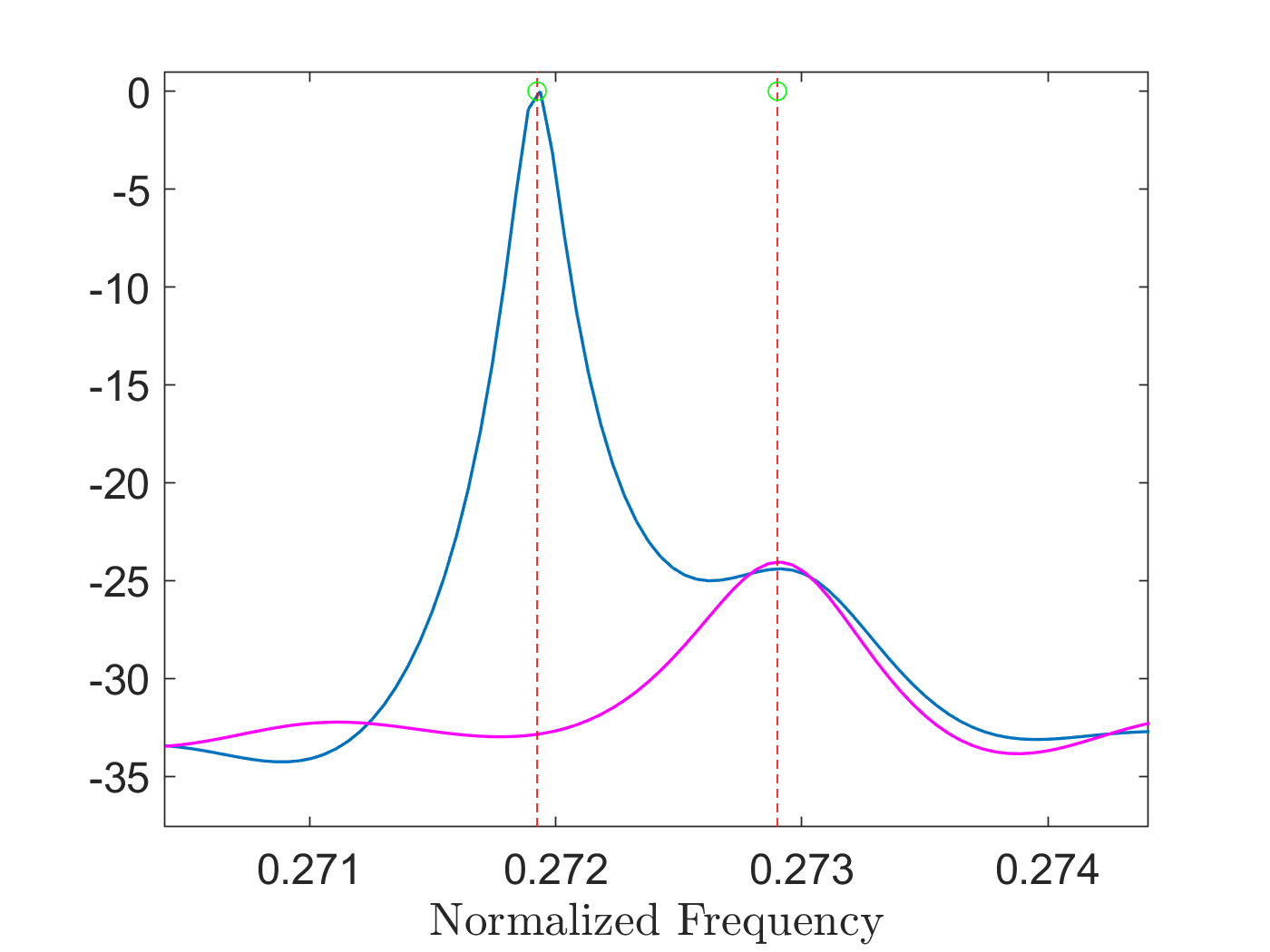}
    \caption{}
    \end{subfigure}
    \caption{\small \sl Comparison of the of MUSIC with trivial peak-picking (a) and peak-picking after nulling (b) for a two source signal with a dynamic range disparity. The true source center frequency is indicated by the dashed red line and the estimated center frequency is given by the green circles. The signal after the main source is nulled out is provided by the magenta line.}
    \label{fig:MUSIC_OMP_Comparison}
\end{figure}

Producing a fair comparison between classical methods such as the MUSIC algorithm and the modern SLE methods is difficult. The sparse sparse algorithms are inherently one-shot (requiring a single set of samples) while MUSIC and similar covarience based methods hinge on having access to multiple sampled realizations of \eqref{eq:source_loc_cont}. Therefore we settle on a qualitative example that demonstrates the utility of a more modern approach to SLE. We consider the case of two signals closely spaced in frequency with appreciable dynamic range disparities. We observe $N=2^{10}$ samples and $P=8$ realizations of the signal to form the data matrix $\mY^{(0)} = [\vy_1 \ \vy_2 \ \dots \vy_P]$. Figure \ref{fig:MUSIC_OMP_Comparison}(a) shows the MUSIC pseudo-spectrum formed from this scenario, and as is apparent the peak location of the smaller source is somewhat ambiguous. Simply using MUSIC and peak picking results in a very poor center frequency estimate due to high sidelobe energy of the large source. However, if we estimate the larger frequency $f_1$ and form a nulled residual $\mY^{(1)} = \left(\mtx{I}-\frac{1}{N}\vphi(f_1)\vphi^H (f_1)\right)\mY^{(0)}$ where $\vphi(f) = [e^{j 2 \pi f \cdot 0},e^{j 2 \pi f \cdot 1},\dots,e^{j 2 \pi f \cdot (N-1)}]$ we can produce a new pseudo-spectrum from $\mY^{(1)}$ in a manner similar to OMP. This is shown in Figure \ref{fig:MUSIC_OMP_Comparison}(b) as a purple line. This nulling step also nulls out the sidelobe energy of the larger source, hence the smaller peak becomes far more visible and results in a far better frequency estimate. Ultimately we can conclude that though these classical methods are effective under in their given model space there is still an enormous benefit to investigating more modern methods of SLE. Finally, we note that additional experiments directly comparing classic and modern SLE methods in a variety of scenarios can be found in \cite{DUARTE2013111,Liao:2014}.

% As will be discussed in the next section, all of the above algorithms fail to generalize when the sources are not narrowband. Furthermore, modifying most of these algorithms to account for bandwidth is in general difficult. In particular, for atomic norm denoising it is not even clear how to define an atomic set for general broadband signals\footnote{Even in the discrete case e.g. $\ell_1$-minimization we must carefully devise some notion of group sparsity.}. Even if such a set existed it is not clear how it could be cast to a tractable framework \cite{Suliman:2021}. On the other hand, the iterative OMP and CoSaMP algorithms can be \textit{modified} to account for bandwidth in a fairly natural manner.   
\section{Numerical Experiments: Broadband Spectral Estimation}
\label{sec:simulation_results}

One of the key findings in our experimental evaluation of techniques for SLE is that greedy algorithms such as OMP and CoSaMP achieve a generally desirable balance between being competitive with the most accurate methods while being also computationally scalable. In light of this, as we move forward we will focus exclusively on similar greedy algorithms, but now with a focus on broadband spectral estimation. Below we offer a series of numerical experiments that compare the performance of our proposed algorithms that exploit Slepian models to more traditional Fourier models used in narrowband SLE. In particular, we observe how both approaches perform when subject to a variety of SNRs and dynamic ranges both with and without subsampling. In line with our previous discussions, the Slepian models enable substantial performance gains compared to simpler Fourier models, especially in more adverse scenarios.

\subsection{Experimental setup}
In the experiments below we set  $N=2^{14}$ and generate signals based on a sum of $\widetilde K$ sinusoids model:
\begin{align}
    \label{eq:sum_of_sines}
    \widetilde\vx_\ell = \sum_{i=1}^{\widetilde{K}}\beta_{i}\vphi(f_i),~\vx_\ell = d_\ell \frac{\widetilde\vx_\ell}{\norm{\widetilde\vx_\ell}_2},
\end{align}
where $f_i \in [f_\ell-W_\ell,f_\ell+W_\ell]$, $\beta_i \sim \mathcal{N}_{\mathcal{C}}(0,1)$, and $d_\ell>0$ is a scale factor. Each $f_{\ell}$ is drawn uniformly at random from the range $[0,1]$ subject to a separation constraint
\begin{align}
    \min_{\ell \neq \ell'}\, |f_{\ell} - f_{\ell'}| \geq  \max_{\ell}\, 4 W_{\ell},
    \label{eq:min_sep}
\end{align}
which ensures no ambiguity/overlap between sources.  We let $\veta \sim \mathcal{N}_{\mathcal{C}}(\mtx{0},\sigma^2\mtx{I})$ with $\sigma^2$ determined by the in-band (channel) SNR. For a fixed $L$ we give each source a different  bandwidth while maintaining a constant occupied bandwidth of $2W_{\text{tot}}$. In cases where the sensing matrix $\mtx{A}$ has $M<N$ we define the oversampling factor
\begin{align}
    \mbox{Oversampling Factor} = \frac{M}{2NW_{\text{tot}}},
\end{align}
to quantify the amount of subsampling. Essentially, this factor compares the number of samples taken to the ideal case where $M$ equals the approximate total degrees of freedom in superposition of signals. In terms of detection, a signal is considered to be correctly estimated when its center frequency estimate $\widehat{f}_{\ell}$ satisfies
\begin{align}
    |f_{\ell} - \widehat{f}_{\ell}| \leq W_{\ell}.
    \label{eq:det_criterea}
\end{align}
For a set of true center frequencies $f_{\ell} \in \{ f_1,\dots,f_L\}$ and estimated center frequencies $\widehat{f}_{\ell} \in \{ \widehat{f}_1,\dots,\widehat{f}_L\}$ we consider the error metric  
\begin{align}
    \text{err}_{f_c} = \frac{1}{L} \sum_{\ell = 1}^L \min{\left(\min_{i=1,\dots,L}|f_{\ell}-\widehat{f}_i|,W_{\ell} \right)}.
    \label{eq:fc_err}
\end{align}
Given a set of estimated bandwidths $\widehat{W}_{\ell} \in \{ \widehat{W}_1,\dots,\widehat{W}_L\}$ the nearest true center frequency and bandwidth index is given by
\begin{align*}
    \widetilde{\ell}(f_{\ell}) = \argmin_{i=1,\dots,L} |f_{\ell}-\widehat{f}_i|.
\end{align*}
We then define the bandwidth error metric
\begin{align}
    \text{err}_{W} = \frac{1}{L} \sum_{\ell = 1}^L \frac{|\widehat{W}_{\widetilde{\ell}(f_{\ell})}- W_{\ell}|}{W_{\ell}}\mathbbm{1}\{ |f_{\ell}-\widehat{f}_{\widetilde{\ell}(f_{\ell})}|\leq W_{\ell} \}
    \label{eq:W_err}
\end{align}
where $\mathbbm{1}\{\cdot\}$ is the indicator function. This ensures that the ``missed" center frequencies are not included in the error calculation. Since the signals are generated stochastically we average the results of each scenario over 50 trials.

\subsection{1-D detection and resolution}
\begin{figure}[t]
        \centering
        \begin{subfigure}[b]{0.4\textwidth}
            \centering
            \includegraphics[width=\textwidth]{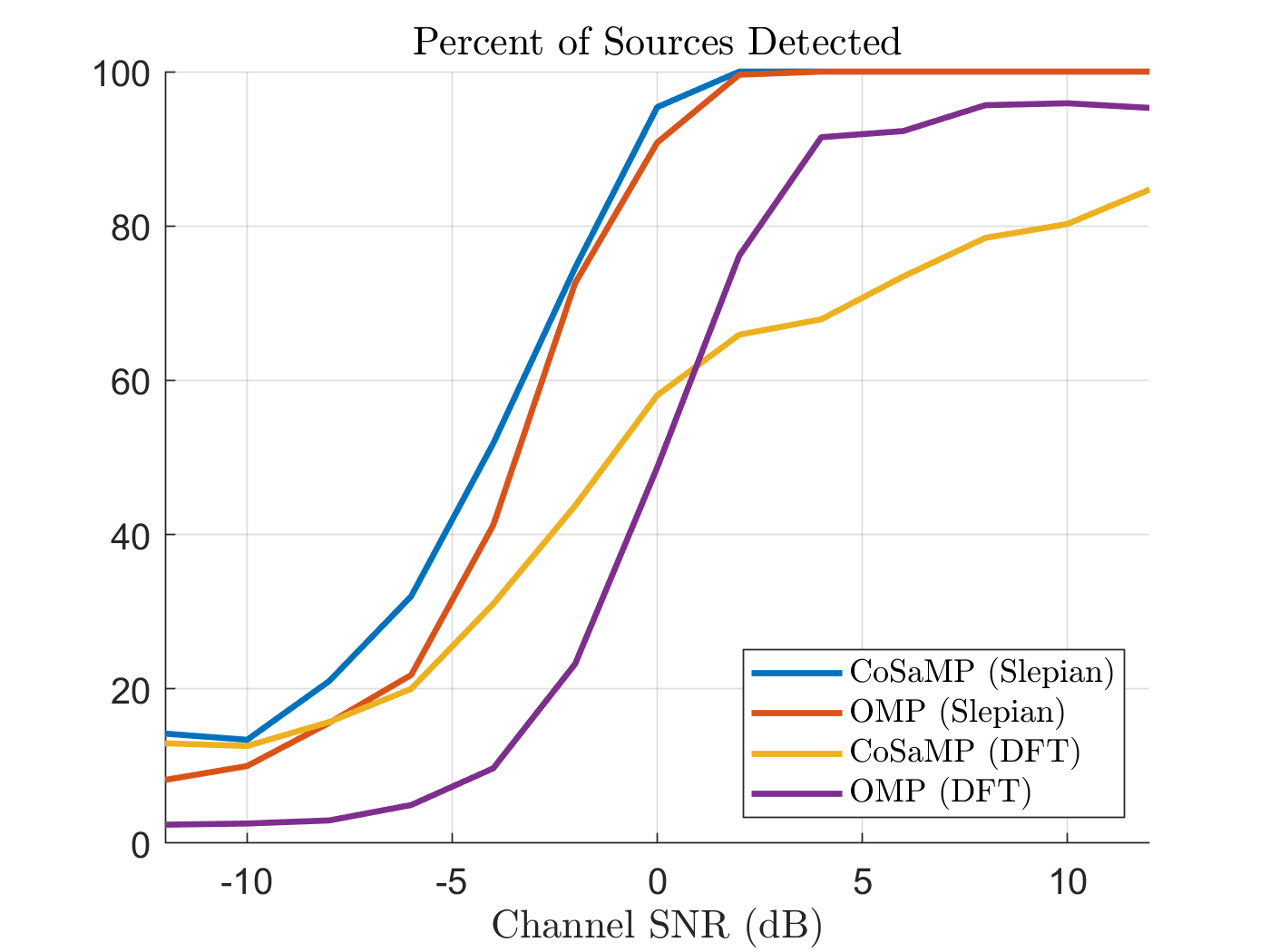}
            \caption[CoSaMP Percent Detected]%
            {{\small}}    
            \label{fig:CoSaMP_perc_Detect_no_sub}
        \end{subfigure}
        \hfil
        \begin{subfigure}[b]{0.4\textwidth}  
            \centering 
            \includegraphics[width=\textwidth]{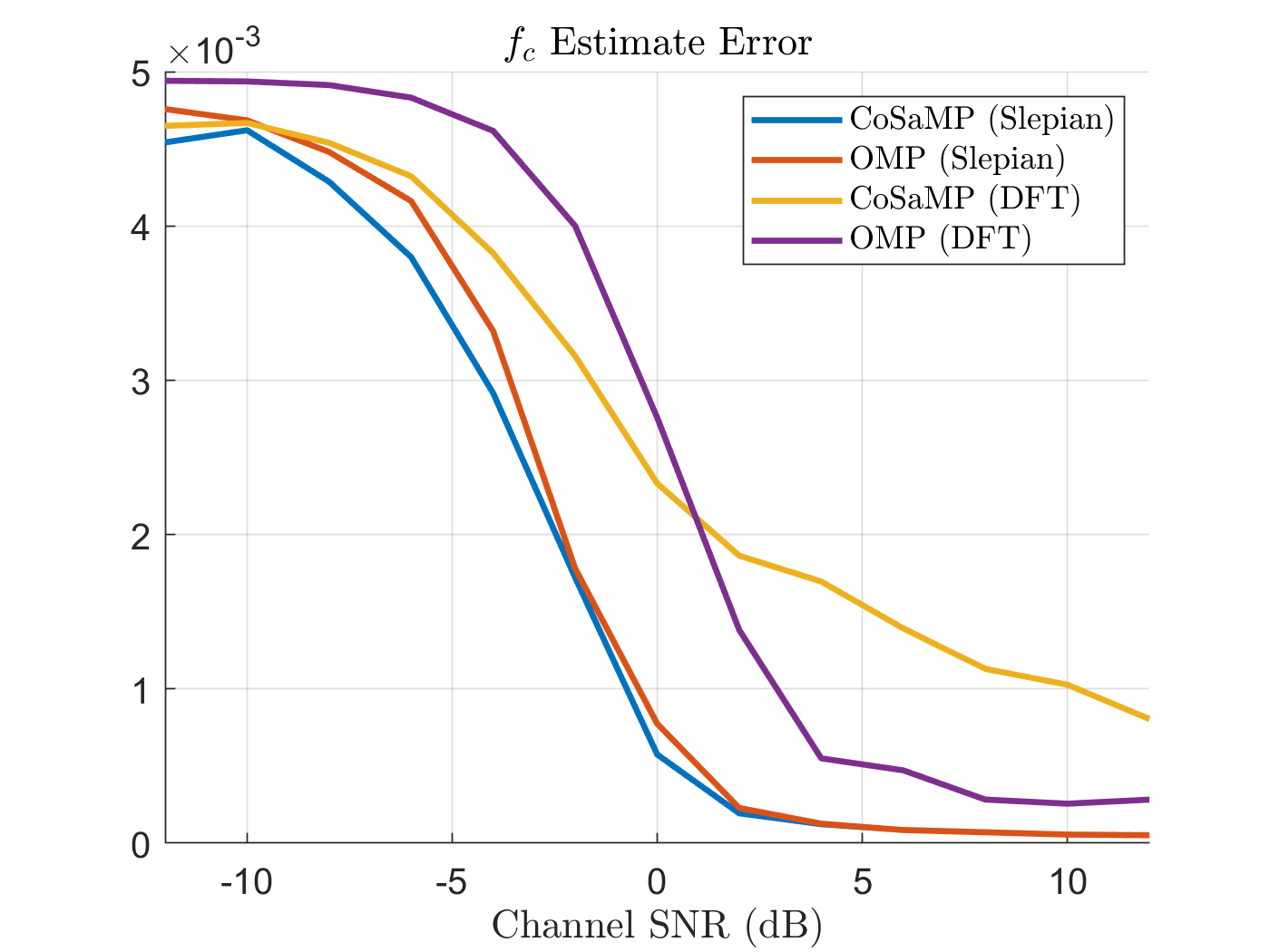}
            \caption[CoSaMP center frequency error]%
            {{\small}}    
            \label{fig:CoSamp_fc_no_sub}
        \end{subfigure}
        \caption[CoSaMP Full sample Performance]
        {\small \sl Performance of CoSaMP and OMP algorithms with fully sampled source when using a DFT or Slepian dictionary. (a) Percent of sources detected. (b) Error in center frequency estimates in normalized frequency units.} 
        \label{fig:Alg_Comparison_v_SNR}
\end{figure}

To begin, we compare algorithms based on the percentage of sources they correctly detect and the relative estimation error under various levels of noise and subsampling. We let $L=10$ and set $2W_{\text{tot}} = 0.04$ and fix the dynamic range such that $d_{\ell} = 1$ for $\ell = 1,\dots,L$. Additionally, we examine how generating $\mPsi$ using a Slepian basis compares to a Fourier dictionary for $\mPsi$. The results in Figure~\ref{fig:Alg_Comparison_v_SNR} show the detection and resolution metrics for the CoSaMP and OMP algorithm when using either a Fourier or Slepian $\mPsi$. It is clear from these results that the Slepian $\mPsi$ yields far better detection performance across a wider variety of channel SNRs, as expected. As a further observation, when using a Slepian dictionary, CoSaMP seems to slightly outperform OMP in both detection and resolution. However, this may simply be due to a sub-optimal choice in the OMP hyperparmeters -- the performance gap could possibly be narrowed by fine tuning of these parameters.

\begin{figure}[t]
        \centering
        \begin{subfigure}[b]{0.4\textwidth}
            \centering
            \includegraphics[width=\textwidth]{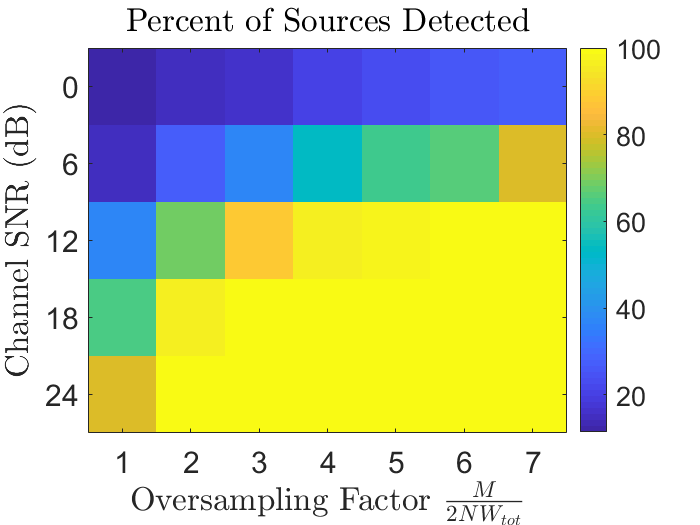}
            \caption[CoSaMP Slepian Percent Detected]%
            {{\small}}    
            \label{fig:CoSaMP_Slepian_Detect}
        \end{subfigure}
        \hfil
        \begin{subfigure}[b]{0.4\textwidth}  
            \centering 
            \includegraphics[width=\textwidth]{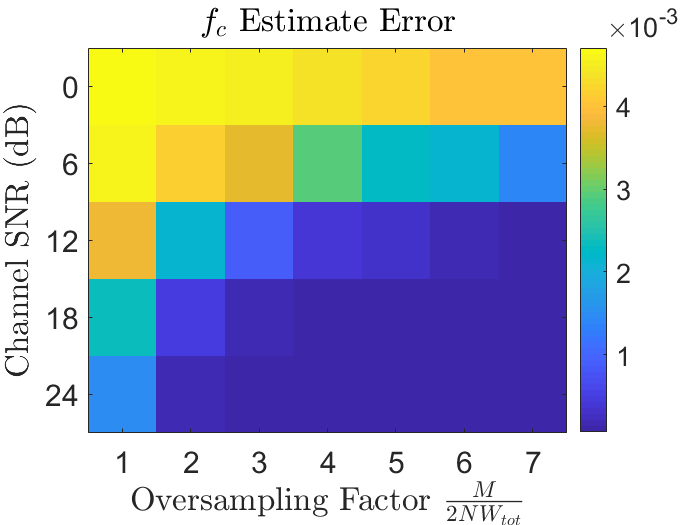}
            \caption[CoSaMP Slepian center frequency error]%
            {{\small}}    
            \label{fig:CoSamp_Slepian_fc}
        \end{subfigure}
        \caption[CoSaMP Subsampling Performance]
        {\small \sl Slepian dictionary based CoSaMP algorithm's localization performance when subject to subsampling. (a) Percent of sources detected. (b) Error in center frequency estimates in normalized frequency units.} 
        \label{fig:CoSaMP_Compressed_Case}
\end{figure}

For the next set of experiments we let $\mtx{A}$ be a random subsampling matrix such that $\vy$ is now comprised of compressed measurements. As previously discussed, Slepians are more aptly suited for capturing the energy generally lost due to spectral leakage than their DFT counterpart. Ultimately this results in a nulling scheme that more effectively handles the subsampling artifacts. Since it had the best performance in the fully sampled case we examine the performance of CoSaMP (Slepian) when subject to varying levels of subsampling and SNRs. The results shown in Figure \ref{fig:CoSaMP_Compressed_Case} demonstrate that similar performance to the fully sampled case can be achieved as long as the signal is oversampled by a factor greater than 7.

\subsection{Effects of dynamic range}
The next set of experiments are designed to test how well the algorithms perform when the component signals in \eqref{eq:source_loc_comp} have large dynamic range disparities. We choose $L=2$ and fix the total occupied bandwidth to $2W_{\text{tot}} = 0.02$. The dynamic range (DR), given in units of dB, is varied such that $d_{1} = 1$ and $d_{2} = 10^{-\frac{\text{DR}}{20}}$. Signals are generated under a modified separation constraint
\begin{align}
    \min_{\ell \neq \ell'}\, |f_{\ell} - f_{\ell'}| =  \max_{\ell}\, \Delta_{W} W_{\ell},
    \label{eq:min_sep_mod}
\end{align}
and the parameter $\Delta_{W}$ is varied depending on the experiment. We fix the channel SNR to $24$ dB such that we are operating in a regime where it is known the algorithms have good performance in the case where $d_1=d_2$ (a dynamic range of 0dB). Again, we compare algorithms based on the percentage of sources they correctly detect and the relative estimation error. However, we make one minor change to the detection criteria; we consider a correct detection to be when both signals in a trial are detected and then average over the trials. This is to account for the fact that the signal with more energy is almost always detectable. As previously suggested for this scenario, under high dynamic range scenarios we set $\epsilon_c$ to be far smaller than the previous experiments.

\begin{figure}[tbp]
        \centering
        \begin{subfigure}[b]{0.4\textwidth}
            \centering
            \includegraphics[width=\textwidth]{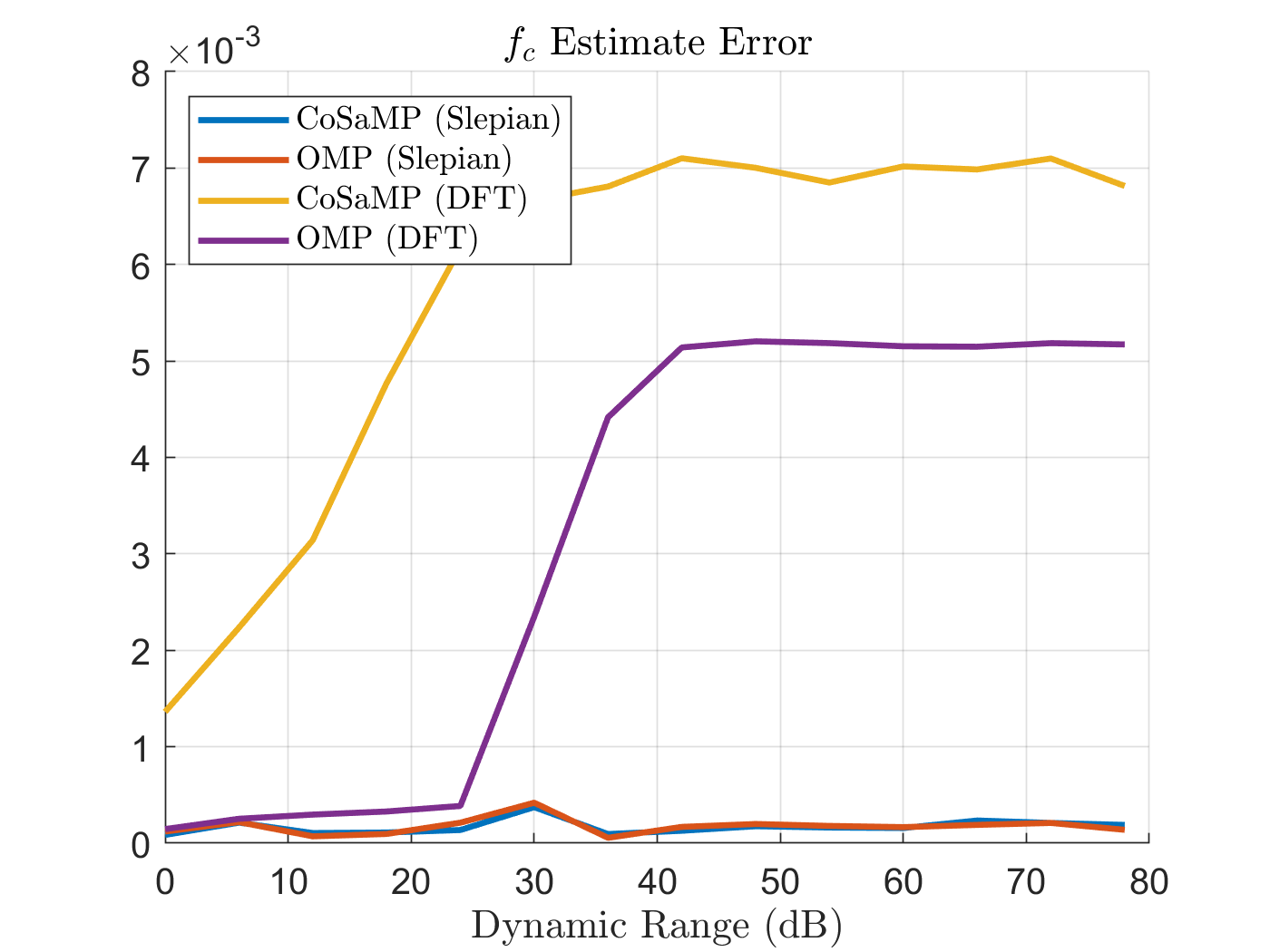}
            \caption[CoSaMP Slepian Percent Detected]%
            {{\small}}    
            \label{fig:DR_Detect}
        \end{subfigure}
        \hfil
        \begin{subfigure}[b]{0.4\textwidth}  
            \centering 
            \includegraphics[width=\textwidth]{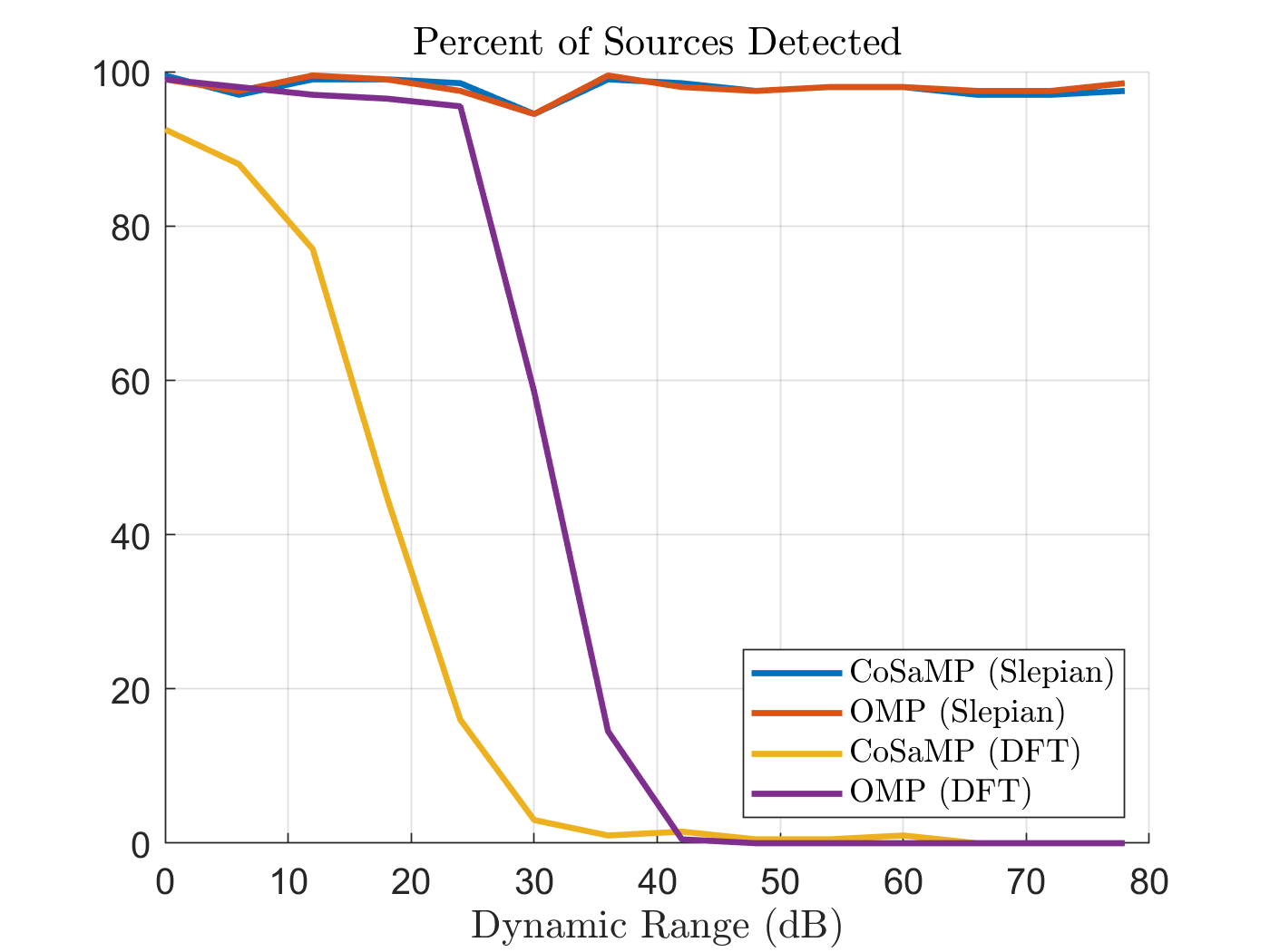}
            \caption[CoSaMP Slepian center frequency error]%
            {{\small}}    
            \label{fig:DR_fc}
        \end{subfigure}
        \caption[CoSaMP Subsampling Performance]
        {\small \sl Performance of CoSaMP and OMP algorithms with fully sampled source when using a DFT or Slepian based dictionary when subject to a variety of dynamic ranges. (a) Percent of sources detected. (b) Error in center frequency estimates in normalized frequency units.} 
        \label{fig:Dynamic_Range_1D}
\end{figure}

We begin by examining the fully sampled case and fix $\Delta_{W} = 4$. The results in Figure \ref{fig:Dynamic_Range_1D} show the detection and error metrics for the CoSaMP and OMP algorithm when using either a Fourier or Slepian $\mPsi$ for a variety of dynamic ranges. It is apparent from these results that the algorithms that use Slepian models for $\mPsi$ are almost agnostic to the dynamic range of the sources within the tested regime. In contrast the DFT dictionary based methods show a performance roll off around $30$ dB. This is largely due to the Slepian dictionary's ability to deal with spectral leakage, allowing for consistent detection and nulling in scenarios where the DFT simply bleeds too much out of band energy to be viable. 

\begin{figure}[tbp]
        \centering
        \begin{subfigure}[b]{0.4\textwidth}
            \centering
            \includegraphics[width=\textwidth]{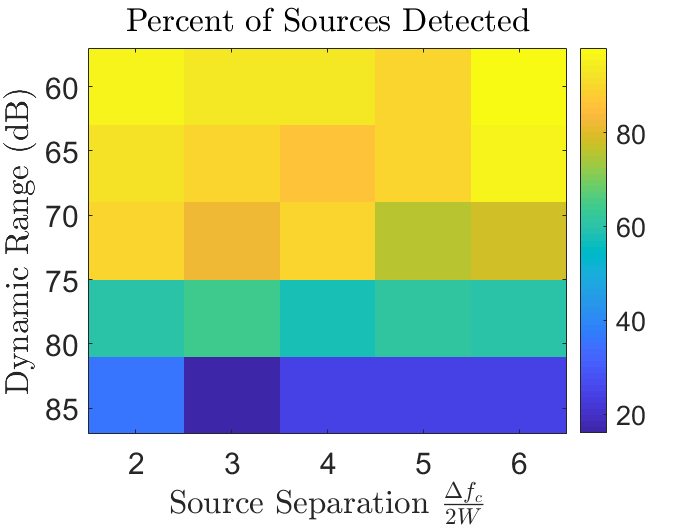}
            \caption[CoSaMP Slepian DR Percent Detected]%
            {{\small}}    
            \label{fig:CoSaMP_Slepian_DR_Detect}
        \end{subfigure}
        \hfil
        \begin{subfigure}[b]{0.4\textwidth}  
            \centering 
            \includegraphics[width=\textwidth]{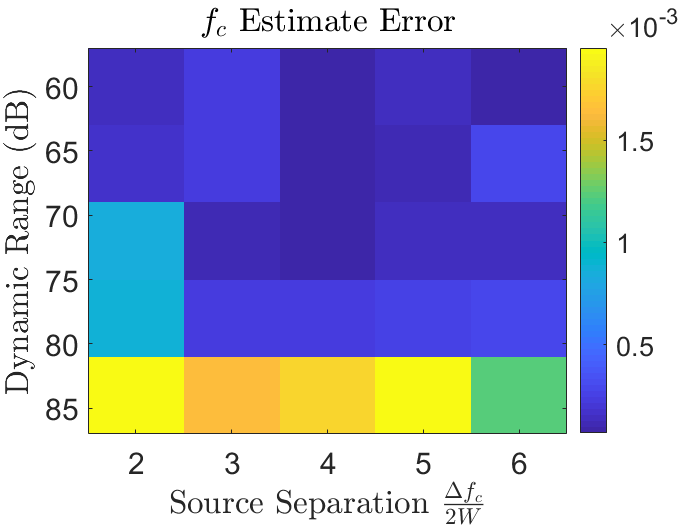}
            \caption[CoSaMP Slepian DR center frequency error]%
            {{\small}}    
            \label{fig:CoSamp_Slepian_DR_fc}
        \end{subfigure}
        \caption[CoSaMP Subsampling Performance]
        {\small \sl Slepian dictionary based CoSaMP algorithm's localization performance when subject to large dynamic ranges and closely spaced sources. (a) Percent of sources detected. (b) Error in center frequency estimates in normalized frequency units.} 
        \label{fig:CoSamp_Slepian_Compressed_DR_comparison}
\end{figure}

For the case where $\mtx{A}$ is a random subsampling matrix we fix the oversampling ratio to 8 and vary $\Delta_{W}$ in order to see the effects of the source spacing, which will cause $\mPsi$ to become increasingly coherent. The results shown in Figure \ref{fig:CoSamp_Slepian_Compressed_DR_comparison} indicate that even with compressed measurements we can have fairly good performance over a wide range of dynamic range disparities. The source separation seems to not impact performance to a large degree, which again is due to the high spectral concentration of the Slepian basis vectors.
\section{Application to Localization in Multi-Sensor Arrays}
\label{sec:multi_sensor_array}

A particularly interesting application of our broadband source localization algorithms is the setting when our signal impinges on a $P$-element\footnote{In the context of multi-sensor arrays we will use $P$ to refer to the number of sensors. This is in contrast to Section~\ref{sec:numerical_SLE} where $P$ was the number of realizations of samples from \eqref{eq:source_loc_cont}.} multi-sensor array as a plane wave. In a sense this formulation couples two common source localization problems; spectral support estimation and DOA. This section provides experiments that highlight how a modest modification of our algorithm can be applied to this scenario. Particular emphasis is placed on the ability to spatially localize broadband signals, which is generally difficult without an underlying narrowband assumption.

\subsection{Extension to space-time array processing}
% In this scenario the signals can be seperated in terms of both their respective temporal frequency support \emph{and} spatial frequency support. The source localization problem then seeks to estimate both of these supports, with the latter often termed DOA. Details on how this changes our model can be found in Appendix~\ref{ap:multi_sensor_array}, but the key point is that the addition of $P$ delayed versions of the signal introduces an immense amount of redundancy to the system. We can leverage these redundancies to localize the signal in space as well as improve our temporal frequency support estimates by raising the SNR through the proverbial ``array gain." By assuming the respective spatial and temporal spaces can be treated separably we can extend Algorithms~\ref{alg:OMP_compressed} and \ref{alg:CoSaMP_compressed} to operating in this regime with minimal modification.

Here we give a brief overview of array processing and how it changes our signal models. To begin we assume there are a series of $P$ sensors placed at positions $\vz_p \in \mathbb{R}^3$. A signal $x_\ell(t)$, traveling through space as a plane wave impinges on this set of sensors at an angle $\vtheta_\ell=[\phi_\ell,~\theta_\ell]^T$ where $\phi_\ell$ and $\theta_\ell$ are the azimuth and elevation angles respectively. The signal that actually reaches the sensors is a delayed version of $x_\ell(t)$:
\begin{align}
    \label{eq:comp_sig_array}
    x_{\ell,p}(t) = x_\ell(t-\tau_{\ell,p}),
\end{align}
where $\tau_{\ell,p}$ depends on the relative position of the $p^{\text{th}}$ sensor to the array's  phase-center and $\vtheta_\ell$. More explicitly, let $\vu_\ell = [\cos\phi_\ell \cos\theta_\ell,~\sin\phi_\ell \cos\theta_\ell,~\sin\theta_\ell]^T$ be a normal vector associated with the $\ell^{\text{th}}$ plane wave relative to the array center and $c$ be the speed of light. Then the delays in \eqref{eq:comp_sig_array} are $\tau_{\ell,p} = \inner{\vu_\ell,\vx_{p}}/c$. Returning to a model similar to \eqref{eq:source_loc_cont} we assume the $p^{\text{th}}$ sensor observes a superposition of $L$ sources
\begin{align}
    \label{eq:full_sig_array}
    y_{p}(t) = \sum_{\ell=1}^L x_{\ell,p}(t) + \eta_{p}(t),
\end{align}
and that we have access to the ensemble of outputs $\{y_{p}(t)\}_{p=1}^{P}$. What is interesting about \eqref{eq:full_sig_array} is that we now have two notions of separation between sources; temporal \emph{and} spatial. For instance, consider the case where $f_1=f_2=\dots=f_L$ such that the standard 1-D source localization problem is ill-posed. If at the same time each $\{\vtheta_\ell\}_{\ell=1}^L$ are distinct then we can still resolve the distinct components using the ensemble of outputs. In effect, we are able to utilize separation in the spatial domain to compensate for each source having identical temporal-frequency support. Alongside determining the temporal-frequency support of \eqref{eq:full_sig_array} multi-sensor array broadband source localization also seeks to estimate the  $\{\vtheta_\ell\}_{\ell=1}^L$ (i.e., the DOA) of the sources from a finite collection of samples taken off the array.

As in the 1-D case we again sample the signal at the Nyquist rate to produce $N$ uniform samples. Collecting each of these $N$ ``snapshots" of the array into the $P\times N$ matrices
\begin{align*}
    \mY = 
    \begin{bmatrix}
    y_1(t_1) & \dots & y_1(t_N)\\
    \vdots & \ddots & \vdots \\
     y_{P}(t_1) & \dots & y_{P}(t_N)
    \end{bmatrix},
    \mX_\ell = 
    \begin{bmatrix}
    x_{\ell,1}(t_1) & \dots & x_{\ell,1}(t_N)\\
    \vdots & \ddots & \vdots \\
     x_{\ell,P}(t_1) & \dots & x_{\ell,P}(t_N)
    \end{bmatrix},
    \mN = 
    \begin{bmatrix}
    \eta_1(t_1) & \dots & \eta_1(t_N)\\
    \vdots & \ddots & \vdots \\
     \eta_{P}(t_1) & \dots & \eta_{P}(t_N)
    \end{bmatrix},
\end{align*}
we express the multi-sensor sampled version of \eqref{eq:source_loc_cont} as
\begin{align}
    \label{eq:array_samp}
    \mY = \sum_{\ell=1}^L \mX_\ell+\mN.
\end{align}
We are also interested in estimating the spectral support when we view the signal through a set of compressed samples. In order to modify \eqref{eq:array_samp} to account for this we consider a set of $P\times N$ sensing matrices $\{\mA_m\}_{m=1}^M$ and let $\mS=\sum_{\ell=1}^L\mX_\ell$. The $m^{\text{th}}$ compressed measurement is then given by 
\begin{align}
    \label{eq:array_comp}
    y_m = \inner{\mA_m,\mS}+\eta_m
\end{align}
where $\inner{\cdot,\cdot}$ denots the Frobenius inner product.\footnote{$\inner{\mA,\mB}=\trace{\mA^H\mB}=\sum_{i,j}\mA_{i,j}^*\mB_{i,j}$.} Given the full collection of $M$ measurement of the form in \eqref{eq:array_comp} we seek to estimate the same parameters as in the fully sampled case of \eqref{eq:array_samp}. 

In order to modify our algorithms to operate in the multi-sensor regime we make one major structural assumption; the spatial and temporal domains can be treated separably. What this allows us to do is perform the subspace identification step separably by first projecting on the candidate temporal-frequency subspaces and then projecting onto the candidate spatial-frequency subspaces. This makes the fast methods described in \cite{FST,Karnik:2022} amendable to the multi-sensor array case. However, it should be noted that the separability assumption is not accurate in the general case as the spatial and temporal domains are inherently coupled, and this coupling becomes more pronounced as the bandwidth of the component signals increases. Though this simplifying assumption buys us the ability to apply fast algorithms, it comes at a cost of producing a sparse representation of \eqref{eq:array_samp} that overestimates the implicit degrees of freedom. To make this more explicit consider the $\ell^{\text{th}}$ component signal in \eqref{eq:array_samp}. We can estimate the spatial and temporal subspace dimensions of $\mX_\ell$ to be $K_s$ and $K_t$ respectively such that its low-dimensional representation has a total of $K_s\cdot K_t$ degrees of freedom. However, it has been recently shown in \cite{DeLude:2022} that the degrees of freedom in $\mX_\ell$ scale more like $\approx K_s+K_t$ meaning that separably modeling the system induces a substantial amount of redundancy. Development of non-seperable fast methods of projecting onto the candidate subspaces remains an open problem for future work.

\subsection{Multi-sensor array numerical experiments}
We assume a uniform linear array (ULA) configuration in which the elements are spaced by $\lambda_s/2$ where $\lambda_s=c/f_s$. Under this assumption the delay the $\ell^{\text{th}}$ source sees at the $p^{\text{th}}$ sensor is $\tau_{\ell,p} = p \frac{f_\ell \lambda_s \cos \theta_{\ell}}{2c}$. For our experiments we set $N = 2^{10}$, $L=5$, and fix the total occupied bandwidth of $2W_{\text{tot}} = 0.05$ with the component signals generated from \eqref{eq:sum_of_sines}. These five components are incident to a $P=32$ element ULA and arrive at angles drawn uniformly at random from the range $[0,\pi]$. We set $d_{\ell} = 1$ for $\ell = 1,\dots,L$ such that there is no dynamic range disparity between sources. As in the previous experiments we vary the SNR and the sub-sampling levels and average the results of each scenario over 50 trials.

\begin{figure}[h]
        \centering
        \begin{subfigure}[b]{0.4\textwidth}
            \centering
            \includegraphics[width=\textwidth]{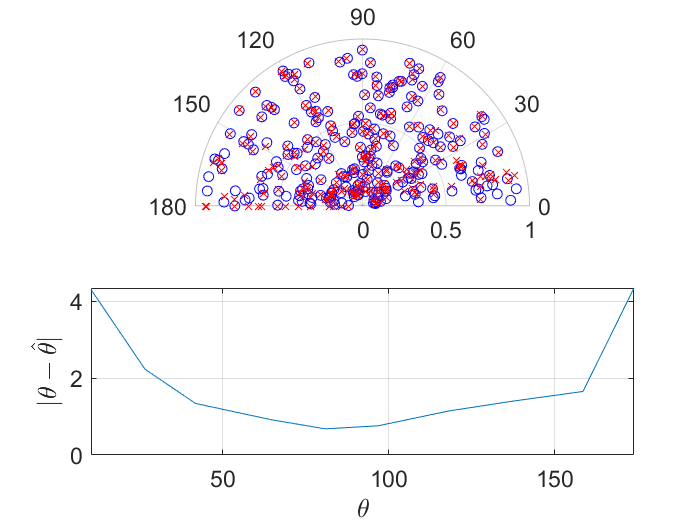}
            \caption[Polar Plot of Source Detection]%
            {{\small}}    
            \label{fig:Polar_Plot_Source_detection}
        \end{subfigure}
        \hfil
        \begin{subfigure}[b]{0.4\textwidth}  
            \centering 
            \includegraphics[width=\textwidth]{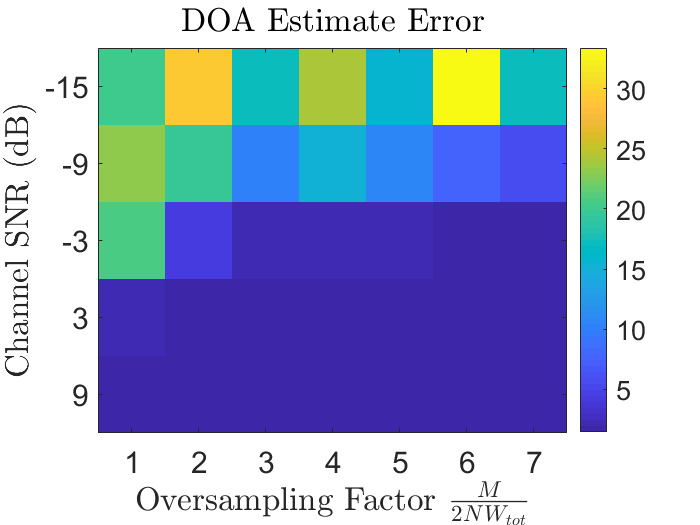}
            \caption[OMP Angular Resolution]%
            {{\small}}    
            \label{fig:OMP_Angular_res}
        \end{subfigure}
        \caption[OMP Array Results]
        {\small \sl Performance of Slepian based OMP algorithm adapted to multi-sensor array scenario. (a) Polar plot of estimated and true sources for an SNR of $3$ dB and an oversampling factor of 7 where $o$ and $x$ denote a source and estimate respectively. The radius is the normalized frequency and the angle is the DOA. Accompanying this is a plot of DOA estimate error in degrees vs the true DOA. (b) Error of DOA estimate in degrees when subject to a variety of channel SNR and subsampling levels.} 
        \label{fig:OMP_Angular_results}
\end{figure}

For both OMP and CoSaMP the multi-sensor array data offers an immense boost in performance when operating in a low SNR regime. This is due to the coherent array gain producing a $10\log_{10}(32)\approx 15.05$ dB increase in the ambient SNR. Consequently the results in terms of temporal frequency support estimation follow nearly identical trends to Figure \ref{fig:CoSaMP_Compressed_Case} but at lower SNRs. Since there is no notable change in trend other than the SNR regime we concentrate on presenting results related to DOA estimation. Figure \ref{fig:OMP_Angular_results} depicts the DOA performance when subject to varying levels of SNR and subsampling. It is clear that when operating at a sufficient SNR and subsampling level the DOA estimates are well within the expected resolution\footnote{For a set of $P$ uniform samples taken at a sampling rate $f_s$ the expected frequency resolution is $f_s/P$.} of $2\pi/P$. However, the results exhibit an angular error dependence with sources close to the endfire possessing a higher estimation error compared to those signals incident at less extreme angles.
\section{Conclusion}
\label{sec:conclusion}
In this paper we have presented a thorough review of existing narrowband source localization algorithms, ranging from classical methods to more advanced optimization based modern techniques. Building on these narrowband methods, we then developed two novel broadband source localization algorithms that were demonstrated to successfully localize signals with arbitrary bandwidth. Furthermore, these algorithms are shown to perform well even when given a set of compressed measurements. The broadband Slepian space model coupled with MTSE was demonstrated to be far more robust to dynamic range disparities and noise than the standard Fourier based modeling methods. Finally, we extended these principles, originally formulated in the 1-D case, to operate on multi-sensor arrays. Future work in this direction could center on the analysis of these algorithms as well as investigating efficient methods of implementing non-seperable models in the multi-sensor array case.

%\section{Introduction to broadband DOA and channel estimation}

%\subsection{Broadband signal model}
%\subsection{Problem setup}

%Broadband DOA estimation - compare with sparse spectral estimation problem 

%\section{Lit survey on broadband DOA estimation}

%\input{algorithms}
%\input{experiments}

\bibliographystyle{unsrt}
\bibliography{IEEEabrv.bib, ref}
%\appendix
%\include{Sections/Appendix_CoSaMP_OMP}
%\include{Sections/Appendix_Array_Detection}

\end{document}